\documentclass[a4paper,11pt]{article}
\pdfoutput=1 
\usepackage{jheppub} 
\usepackage{bbm,bm,graphicx,mathtools,color,hyperref,slashed}
\usepackage{amssymb,amsmath}
\usepackage[T1]{fontenc}

\usepackage[all]{xy}

\graphicspath{{./Figures/}}

\newcommand{\diff}{\mathrm{d}}

\newcommand{\ve}{\varepsilon}
\newcommand{\Diff}{{\mathcal{D}}}

\newcommand{\be}{\begin{equation}}      
\newcommand{\ee}{\end{equation}}      
\newcommand{\bea}{\begin{eqnarray}}      
\newcommand{\eea}{\end{eqnarray}}

\newcommand{\tr}{\mathrm{tr}}

\newcommand{\im}{\mathrm{i}}

\newcommand{\rme}{\mathrm{e}}

\preprint{YITP-20-45}

\title{Universality between vector-like and chiral quiver   gauge theories: Anomalies and domain walls
}

\author[a]{Tin Sulejmanpasic,}

\affiliation[a]{Department of Mathematical Sciences, Durham University, Durham, DH1 3LE, United Kingdom}

\author[b,c]{Yuya Tanizaki,}

\affiliation[b]{Yukawa Institute for Theoretical Physics, Kyoto University, Kyoto 606-8502, Japan}

\author[c]{Mithat \"{U}nsal}

\affiliation[c]{Department of Physics, North Carolina State University, Raleigh, NC 27695, USA}

\emailAdd{tin.sulejmanpasic@durham.ac.uk}
\emailAdd{yuya.tanizaki@yukawa.kyoto-u.ac.jp}
\emailAdd{unsal.mithat@gmail.com}

\abstract{
We study low-energy dynamics of $[SU(N)]^K$ chiral quiver gauge theories in connection with $\mathcal{N}=1$ super Yang-Mills (SYM) theory,  and  quantum chromodynamics with bi-fundamental fermions (QCD(BF)). 
These theories can be obtained by $\mathbb{Z}_K$ orbifold projections of $\mathcal{N}=1$ $SU(NK)$ SYM theory, but the perturbative planar equivalence does not extend nonperturbatively for $K\ge 3$.   
In order to study low-energy behaviors, we analyze these systems using 't~Hooft anomaly matching and reliable semiclassics on $\mathbb{R}^3\times S^1$. 
Thanks to 't~Hooft anomaly that involves $1$-form center symmetry and discrete chiral symmetry, we predict that chiral symmetry must be spontaneously broken in the confinement phase, and there exist $N$ vacua.  
Theories with even $K$ possess a physical $\theta$ angle despite the presence of massless fermions, and we further predict the $N$-branch structure associated with it; the number of vacua is enhanced to $2N$ at $\theta=\pi$ due to spontaneous $CP$ breaking. 
Both of these predictions are explicitly confirmed by reliable semiclassics on $\mathbb{R}^3\times S^1$ with the double-trace deformation. 
Symmetry and anomaly of  odd-$K$  theories are the same as those of the ${\cal N}=1$ SYM, and the ones of even-$K$ theories are same as those of QCD(BF). 
We unveil why there exists universality between vector-like and chiral quiver theories, and conjecture that their ground states can be continuously deformed without quantum phase transitions. We briefly discuss anomaly inflow on the domain walls connecting the vacua of the theory and possible anomaly matching scenarios.
}

\begin{document}
\maketitle
\section{Introduction and Summary}\label{sec:introduction}

\begin{figure}[t]
\vspace{-1.5cm}
\begin{center}
\includegraphics[width = 0.7\textwidth]{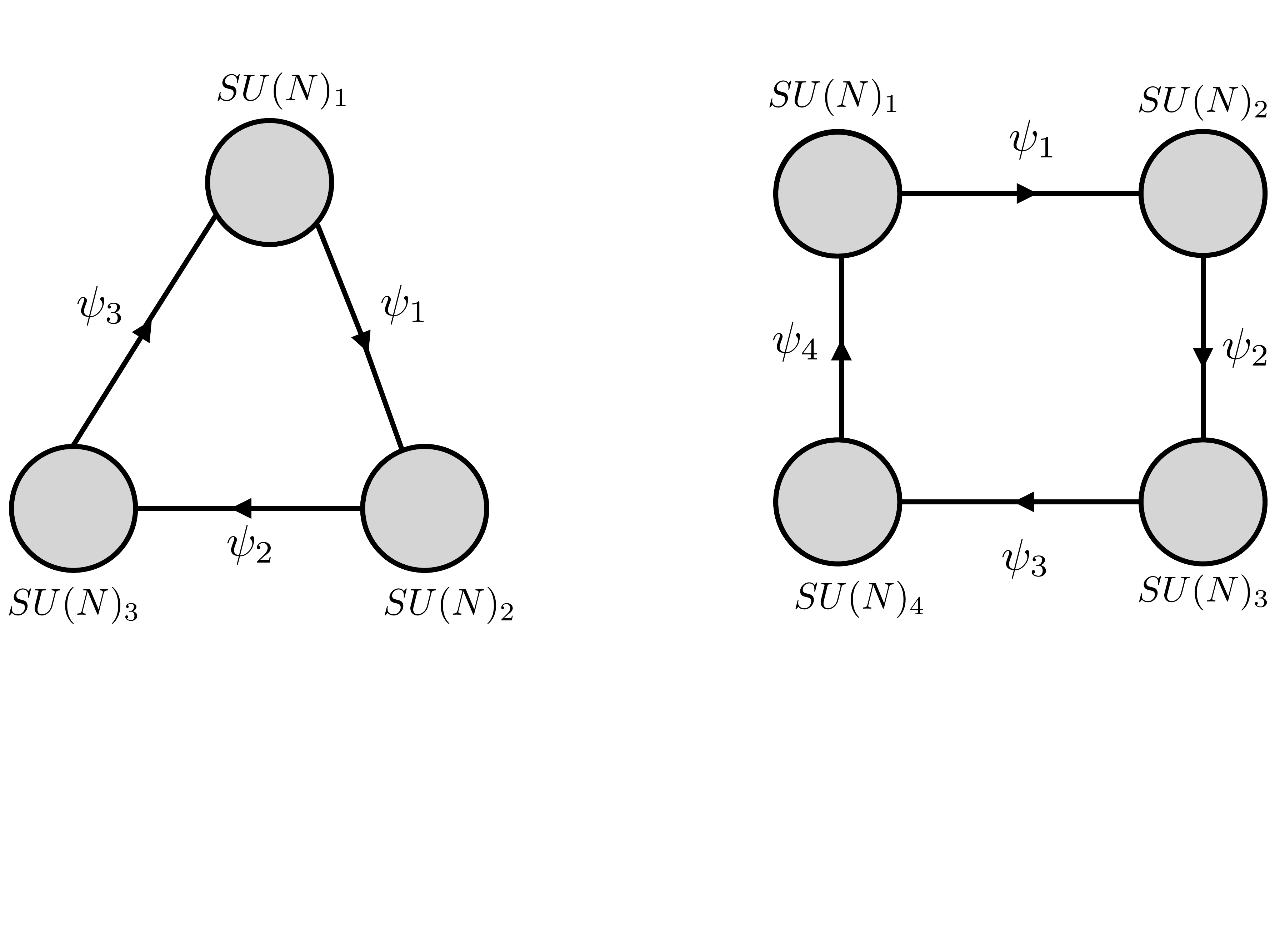}
\vspace{-2cm}
\caption{Simplest examples of chiral quiver theories with odd  $(K=3)$ and even  $(K=4)$ number of sites. The symmetries, symmetry  realizations  and anomalies of the $K$ odd  quiver theories are same as vector-like ${\cal N}=1$ SYM, and the one of the $K$ even theories are same as vector-like QCD with bi-fundamental fermions. 
 $K$ even theories  have irremovable $\theta$ angle despite the existence of massless fermions, which alters the dynamics significantly.
}
\label{fig:quiver}
\end{center}
\end{figure}

In this paper, we will uncover the low-energy dynamics of  chiral quiver gauge theories with 
 $[SU(N)]^K$ gauge group. These theories turns out to have   intriguing relations to vector-like ${\cal N}=1$ supersymmetric Yang-Mills (SYM)  theory  and  
 vector-like  $[SU(N)]^2 $  quantum chromodynamics (QCD) with bi-fundamental fermions (QCD(BF)).  
The matter content of these  theories is   Weyl fermions, $\psi_1,\ldots, \psi_K$, coupled in bifundamental representations of consecutive gauge group factors, forming a quiver graph  with $K$ nodes (See Fig.~\ref{fig:quiver}). 
When $K\ge 3$, a mass term for fermions is forbidden. Therefore,  they provide a class of strongly-coupled chiral gauge theories, and  studying their low-energy properties is a difficult task.

A useful fact about these vector-like ($K=1,2$) and chiral ($K\ge3$) theories is the following.   They can be obtained from $\mathcal{N}=1$ $SU(NK)$ super Yang-Mills (SYM) theory by $\mathbb{Z}_K$ orbifold projection~\cite{Douglas:1996sw}, and thus these theories share the same perturbative expansion in 't~Hooft coupling thanks to the perturbative planar equivalence~\cite{Kachru:1998ys, Bershadsky:1998cb}. 
The perturbative planar equivalence already implies that $\beta$ functions 
and strong scales of these theories are exactly the same at the leading order in large-$N$, as it can be checked by explicit computation.
It is then quite natural to ask if such orbifold equivalence can be true also at the nonperturbative level~\cite{Schmaltz:1998bg, Strassler:2001fs}. If so, we could have understood the low-energy dynamics of this chiral gauge theory by using the nonperturbative knowledge of $\mathcal{N}=1$ SYM. 
Unfortunately, the nonperturbative orbifold equivalence does not hold for  this class of theories if $K\ge 3$~\cite{Kovtun:2004bz, Kovtun:2005kh, Kovtun:2007py}. 
The necessary and sufficient condition for the nonperturbative equivalence requires the absence of spontaneous symmetry breaking of  certain global symmetries  in both parent and daughter theories.  In the parent side, the orbifold projection symmetry must be unbroken, and in the daughter side, the cyclic permutation symmetry between gauge groups must be unbroken. 
In the orbifold projection, we use the subgroup $\mathbb{Z}_K$ of $\mathbb{Z}_{2NK}$ discrete chiral symmetry of $\mathcal{N}=1$ SYM, and this $\mathbb{Z}_K$ subgroup is spontaneously broken if $K\ge 3$\footnote{For $K=2$, the parent theory does not have a problem for nonperturbative orbifold equivalence, because $\mathbb{Z}_2$ is the fermion parity and cannot be spontaneously broken under the assumption of Lorentz-invariant vacuum. So, the validity of nonperturbative equivalence depends on unknown strong dynamics of the daughter side, QCD(BF), which is not completely settled yet~\cite{Kovtun:2005kh, Kovtun:2007py, Armoni:2005wta}.  But using center-stabilizing double-trace deformation  at small $\mathbb R^3 \times S^1$,  one can show that the $\mathbb Z_2$ shift symmetry is unbroken in the semi-classical regime, and all neutral sector observables agree between ${\cal N}=1$  SYM and QCD(BF) \cite{Shifman:2008ja}. 
Throughout this paper, we assume the simplest dynamics for QCD(BF), where the nonperturbative orbifold equivalence works for $K=2$. }. 
Because of this problem on the parent side, planar equivalence does not give useful information on the low-energy dynamics of the daughter chiral theory.

This necessitates new techniques to understand such chiral gauge theories. In this paper, we take an approach from two perspectives: One is the recent generalization of 't~Hooft anomaly matching, and the another is the reliable semiclassics on $\mathbb{R}^3\times S^1$ with double-trace deformation. 
By combining the results of these analyses, we will argue that the chiral symmetry breaking is spontaneously broken,  and that there are $N$ different vacua (except  for $\theta=\pi$ in the $K$ even theory where the vacua doubles  as we shall also discuss).

When a global symmetry $G$ exists in quantum field theory (QFT), we can study the partition function, $\mathcal{Z}[A]$, under the existence of $G$-background gauge field $A$. If such gauged partition function violates the gauge invariance as $\mathcal{Z}[A+\delta_\xi A]=\exp(\im \mathcal{A}[\xi,A])\mathcal{Z}[A]$, this anomalous violation $\mathcal{A}$ is called an 't~Hooft anomaly.  
Anomaly matching claims that the 't~Hooft anomaly $\mathcal{A}$ is the renormalization-group invariant (up to local counter terms), and thus the low-energy effective theory must reproduce the same anomaly~\cite{tHooft:1979rat, Frishman:1980dq}. 
Since '80s, the applicability of this technique has been limited to ``continuous'' chiral symmetry in even-dimensional relativistic fermions. 
In recent years, it has been recognized that anomaly matching holds for much wider class of symmetries~\cite{Wen:2013oza, Kapustin:2014lwa, Cho:2014jfa}, 
and many strongly-interacting field theories are now studied in this viewpoint (see, e.g., Refs.~\cite{Witten:2016cio, Tachikawa:2016cha, Gaiotto:2017yup, Tanizaki:2017bam, Kikuchi:2017pcp, Komargodski:2017dmc, Komargodski:2017smk, Shimizu:2017asf, Wang:2017loc,Gaiotto:2017tne, Tanizaki:2017qhf, Tanizaki:2017mtm,  Yamazaki:2017dra, Guo:2017xex,  Sulejmanpasic:2018upi, Tanizaki:2018xto, Yao:2018kel, Kobayashi:2018yuk,  Tanizaki:2018wtg, Yamaguchi:2018xse, Anber:2018jdf,   Anber:2018xek,Cordova:2018acb, Armoni:2018bga, Yonekura:2019vyz,   Nishimura:2019umw,Cordova:2019jnf,Cordova:2019uob,  Misumi:2019dwq, Anber:2019nfu, Cherman:2019hbq, Bolognesi:2019fej, Tanizaki:2019rbk}). 

Constraints by 't~Hooft anomalies are an exact statement of QFT and should be viewed as kinematic constraints by symmetry.
To match the anomaly, some nontrivial effective theory must appear at low energies, but usually there are several candidates that are compatible with the anomalies. 
In order to go beyond the constraint by anomaly, we study the reliable semiclassics on $\mathbb{R}^3\times S^1$ with the double-trance deformation~\cite{Unsal:2007jx,Unsal:2007vu,Unsal:2008ch, Poppitz:2011wy, Poppitz:2012sw,Poppitz:2012nz, Argyres:2012vv,Argyres:2012ka, Anber:2011gn, Anber:2015wha}. 
In this setup, we can compute the partition function explicitly by dilute gas approximation of monopole-instantons and bions,  and this computation is free from infrared divergence if the compactification size $L$ satisfies $L\ll 1/(N\Lambda)$, where $\Lambda$ is the strong scale. 
Using the idea of Eguchi-Kawai volume independence~\cite{Eguchi:1982nm, GonzalezArroyo:1982hz,GonzalezArroyo:1982ub, GonzalezArroyo:2010ss, Kovtun:2007py}, it is expected that the vacuum expectation values of single-trace operators do  not depend on the compactification size $L$ if $L\gtrsim 1/\Lambda$. 
In the case of pure Yang-Mills theory, this volume independence has been checked for topological susceptibility in the numerical lattice simulation~\cite{Bonati:2018rfg}. 
Moreover, we expect that the adiabatic continuity works, which is a milder version of volume independence, and the dynamics obtained in the above semiclassics shows qualitatively the same behavior of the strongly-coupled dynamics on $\mathbb{R}^4$. 
Even though this is a very nontrivial conjecture, it provides a very useful information on dynamics, and it turns out that the constraint by anomaly is indeed satisfied~\cite{Tanizaki:2019rbk} (For related previous studies, see, e.g., Refs.~\cite{Tanizaki:2017qhf, Tanizaki:2017mtm,  Yamazaki:2017dra}). 

As we will see in this paper, the chiral gauge theory of our interest only has the discrete chiral symmetry and does not have continuous chiral symmetry~\cite{Strassler:2001fs}.  
Because of this fact, it is quite essential to use the recent generalization of anomaly matching in order to understand the low-energy physics. 
In Sec.~\ref{sec:quiver}, we first carefully analyze the global symmetry of this system. When $K$ is odd, the internal global symmetry consists of the $\mathbb{Z}_N$ $1$-form symmetry, or the center symmetry, and of the discrete chiral symmetry $\mathbb{Z}_{2N}$ \footnote{When the gauge couplings of the $[SU(N)]^K$ gauge groups are the same,  there is also a cyclic permutation $\mathbb{Z}_K$ symmetry. Its implication will be discussed in detail in Sec.~\ref{sec:permutation}.}:
\be
(\mathbb{Z}_N^{[1]})_{\mathrm{center}}\times (\mathbb{Z}_{2N})_{\mathrm{chiral}}\quad (K:\,\mathrm{odd}). 
\label{odd-1}
\ee
When $K$ is even, the ordinary internal symmetry consists of the discrete chiral symmetry, $\mathbb{Z}_N$ and the vector-like $U(1)/\mathbb{Z}_N$ symmetry, which defines the baryon number: 
\be
(\mathbb{Z}_N^{[1]})_{\mathrm{center}}\times {U(1)\over \mathbb{Z}_N} \times (\mathbb{Z}_N)_{\mathrm{chiral}}\quad (K:\,\mathrm{even}).  
\label{even-1}
\ee
This important difference between odd and even $K$ was already noticed in Ref.~\cite{Strassler:2001fs}.
The symmetry \eqref{odd-1} is identical to the global symmetry of the vector-like ${\cal N}=1$ SYM  theory  and  \eqref{even-1} is identical to the one of QCD with bi-fundamental fermions.

These theories with even $K$ also present a novel phenomenon about the $\theta$ angles. 
When $K$  is even, there exists a {\it physical} 
 $\theta$ angle which cannot be removed by  chiral rotations. This is true  despite the existence of the massless fermions, both in vector-like $K=2$ and chiral  $K \geq 4$ theories. 
On the other hand, the odd-$K$ theories are extremely rigid. They  have no relevant or marginal deformations on $\mathbb R^4$ except the gauge couplings.   Therefore, it would be quite interesting to determine exact anomalies (kinematic constraints) and understand their non-perturbative dynamics related to $\theta$.  
We also note that the possible origin of $CP$-violation in these models is the topological $\theta$ angle. Therefore, theories with odd $K$ always have the $CP$ invariance because we can set $\theta_i=0$. On the other hands, theories with even $K$ has $CP$ invariance only at the special values of $\theta$: $\theta=0, \pi$ mod $2\pi$.

In Sec.~\ref{sec:odd_sites}, we discuss the case $K$ is odd. 
Introducing the $\mathbb{Z}_N$ two-form gauge field $B$ for $(\mathbb{Z}_N^{[1]})_{\mathrm{center}}$, we find that the partition function $\mathcal{Z}[B]$ transforms under the discrete chiral symmetry as 
\be
(\mathbb{Z}_{2N})_{\mathrm{chiral}}: \mathcal{Z}[B]\mapsto \exp\left(-{\im N\over 4\pi}\int B\wedge B\right)\mathcal{Z}[B]. 
\ee
Assuming that the system shows confinement, this anomaly requires the spontaneous chiral symmetry breaking,
\be
(\mathbb{Z}_{2N})_{\mathrm{chiral}}\to \mathbb{Z}_2, 
\ee
which leads to $N$ isolated vacua. Both anomaly polynomial and chiral symmetry breaking pattern are identical to  theory. 

Semiclassics  confirms this picture, in a  similar construction with the  analysis in ${\cal N}=1$ SYM theory on $S^1 \times \mathbb R^3$ \cite{Davies:2000nw}. 
One big difference is that the monopole vertex, $\mathcal{M}_{j,i}\equiv \rme^{-8\pi^2/(g^2 N)} \rme^{\im \bm{\alpha}_i\cdot \bm{\sigma}_{j}(x)}    (\bm{\alpha}_i\cdot {\bm \psi}_{j-1}) (\bm{\alpha}_i\cdot {\bm \psi}_{j})$, of each gauge group $SU(N)_j$ does not show gauge invariance under other gauge groups $SU(N)_{j'=j\pm 1}$ \cite{Shifman:2008cx}.
Still, magnetic bions are gauge invariant, and thus, in small $S^1 \times \mathbb R^3$ regime, the magnetic-bion induced potential admits $N$ vacua with confinement. In each one of these vacua, 
 the gauge-invariant product of monopole events  
$  {\cal {MR}}_i \equiv   \prod_{j=1}^{K} {\cal M}_{j,i}(x)$    play a crucial role.  These events have the correct number of fermionic zero modes in order to develop the 
chiral condensate, 
 just like the monopole-instantons in   ${\cal N}=1$ SYM do.   In chiral quiver theory, simplest chiral condensate is  $\tr(\psi_1\cdots\psi_K \psi_1\cdots\psi_K )$ and we find that, in the semi-classical regime, this produce a condensate 
$\Lambda^{3K} \rme^{{2\pi\im  k \over N}}$, which is dictated by the strong scale $\Lambda$ of the theory.   
Within the leading large-$N$ analysis, the $L$ dependence of the chiral condensate disappears  just like ${\cal N}=1$ SYM theory.

In Sec.~\ref{sec:even_sites}, we discuss the case $K$ is even. In this case, we introduce not only the $\mathbb{Z}_N$ two-form gauge field $B$ for the center symmetry $(\mathbb{Z}_N^{[1]})_\mathrm{center}$ but also the $U(1)/\mathbb{Z}_N$ gauge field $A_\mathrm{B}$. 
We denote the partition function as $\mathcal{Z}_\theta[B,A_\mathrm{B}]$, and $\theta$ is the physical $\theta$ parameter. 
Applying the discrete chiral transformation $(\mathbb{Z}_N)_{\mathrm{chiral}}$, we find the following anomaly, 
\be
(\mathbb{Z}_N)_{\mathrm{chiral}}:\mathcal{Z}_{\theta}[B,A_\mathrm{B}]\mapsto \exp\left(-\im{1 \over 2\pi}\int (N B\wedge B+B\wedge \diff A_\mathrm{B})\right)\mathcal{Z}_{\theta}[B,A_\mathrm{B}]. 
\ee
In addition to this mixed anomaly involving discrete chiral symmetry, the partition function also shows the anomalous behavior under the shift $\theta\to \theta+2\pi$: 
\be
\mathcal{Z}_{\theta+2\pi}[B,A_\mathrm{B}]=\exp\left({\im N\over 4\pi}\int B\wedge B\right)\mathcal{Z}_{\theta}[B,A_\mathrm{B}]. 
\ee
Under the assumption of confinement and smooth large-$N$ limit, we predict that these anomalies are matched by two interesting behaviors. First, the discrete chiral symmetry breaking occurs,
\be
(\mathbb{Z}_N)_{\mathrm{chiral}}\to 1, 
\ee
and there are $N$ isolated vacua. 
Moreover, each vacuum has the $N$-branch structure leading to the $1$st order phase transition at $\theta=\pi$. So, the vacuum degeneracy is $N$ for generic values of $\theta$, but it becomes $2N$ at $\theta=\pi$. All of the above statements concerning anomaly polynomials and symmetry realizations hold verbatim in the vector-like QCD(BF).  

Semiclassics reproduces this picture  in a rather intriguing  way, which involves new non-perturbative effects and saddles.  By using magnetic-bion-induced potential, we actually find  $N^2$ gauge-inequivalent minima at the  {\it second-order} in semi-classical expansion, rather than $N$.  This $N^2$-fold degeneracy gets lifted by a fairly  high order effect in semi-classics,  and we eventually get $N$ vacua, each of which supports $N$ branches.   The key observation is that, in the small  $S^1 \times \mathbb R^3$ domain,  there are two  gauge-invariant monopole events,  
${\cal {MR}}_{i, \rm odd}  = \prod_{j \in {\rm odd}} {\cal M}_{j,i} $ and  ${\cal {MR}}_{i, \rm even}  =  \prod_{j \in {\rm even}}  {\cal M}_{j,i} $, whose fermionic zero-mode structure matches with that of the chiral condensate.  
One of these events has the $\theta$-dependence that cannot be removed by chiral rotation.      Correlated events of these two configuration of the form 
 $[{\cal {MR}}_{i, \rm odd} \overline { {\cal {MR}}} _{i, \rm even}]$, which is $2K$-th order in semiclassics, is the leading contribution to produce the $\theta$ dependence of the ground-state energies.
The  chiral condensate ends up having a  dependence on  vacuum label   $\ell$,  $\theta$ angle,   and branch label $\tilde \ell$, but is independent of the compactification radius $L$ in the large-$N$ limit.   

In Sec.~\ref{sec:universality}, we explain why there exists universality between chiral quiver theories and vector-like theories. 
We show that $K$-site chiral quiver theory can be continuously deformed to $(K-2)$-site theory when $K>2$ by considering the limit where one gauge group is much more strongly coupled than others. 
By applying this process iteratively, we can obtain QCD(BF) starting from any even-$K$ theories, and obtain $\mathcal{N}=1$ SYM from odd-$K$ theories. 
We conjecture that there is no quantum phase transition and they share the same ground-state structures, even without nonperturbative orbifold equivalence.

In Sec.~\ref{sec:domain_wall}, we give some brief comments about domain-wall theories. Specifically we focus on the domain wall connecting the two vacua related by the minimal discrete chiral transformation. 
Because of the presence of anomaly involving the discrete chiral symmetry, such domain walls should support nontrivial dynamics, and we show that one of the minimal scenario, for both odd and even $K$, is deconfinement of any nontrivial $N$-ality Wilson loops. 
For even $K$, we also propose another scenario without total deconfinement, and instead there emerges an excitation with a fractional baryon charge.

To summarize, despite the lack of large-$N$ orbifold equivalence   between vector-like and chiral quiver theories, we  reveal  remarkable connections between them, which are valid at arbitrary $N$.   For all odd $K$ chiral theories,  the global symmetries, their  mixed anomalies and symmetry realizations are   identical with those of vector-like ${\cal N}=1$ SYM,  and for all even $K$ chiral theories, these properties are identical with those of vector-like QCD(BF).  In all cases, the mechanism of confinement in the semi-classical domain on small $S^1 \times \mathbb R^3$ is the magnetic bion mechanism and mass gap for gauge fluctuations are the same. Again, in all cases, the corresponding chiral condensate on small $S^1 \times \mathbb R^3$  is independent of $S^1$ radius at the leading order of the large-$N$ limit, and is controlled by the strong scale $\Lambda$.

\section{$SU(N)$ chiral quiver gauge theories}\label{sec:quiver}

We consider a chiral gauge theory with the gauge group 
\be
[SU(N)]^K=SU(N)_1\times SU(N)_2\times \cdots \times SU(N)_K, 
\ee
and its matter contents are given by Weyl fermions $\psi_j$ ($j=1,\ldots, K$) in the bifundamental representation of $SU(N)_j\times SU(N)_{j+1}$:
\be
\psi_j\sim (1,\ldots, 1, \bm{N}_j, \overline{\bm{N}}_{j+1},1,\ldots, 1). 
\ee
The Lagrangian of this theory is given by 
\bea
S&=& \sum_{j=1}^{K} \int \left\{{1\over g_j^2}\tr(F_j\wedge \star F_j)+{\im \theta_j\over 8\pi^2}\tr(F_j\wedge F_j)+\overline{\psi}_j \overline{\sigma}_{\mu}D_\mu \psi_j\right\}. 
\label{eq:quiver_lagrangian}
\eea
Here, $F_j=\diff a_j+\im\, a_j\wedge a_j$ is the field strength of $j$-th $SU(N)$ gauge field $a_j$, $g_j$ is the Yang-Mills coupling constant and $\theta_j$ is the vacuum angle.  
The covariant derivative on $\psi_j$ is given by 
\be
D \psi_j=\diff \psi_j+\im\, a_j \psi_j-\im\, \psi_j a_{j+1}. 
\ee
Throughout this paper, the label $j$ for gauge and fermion species is understood to be a cyclic variable mod $K$. 

The $SU(N)$ gauge transformations are given by 
\be
a_j\mapsto g^{-1}_j a_j g_j -\im g^{-1}_j (\diff g_j), 
\ee
and 
\be
\psi_j \mapsto g_j^{-1} \psi_j g_{j+1}, 
\ee
where $g_j$ is the $SU(N)$-valued gauge transformations. The Lagrangian (\ref{eq:quiver_lagrangian}) is invariant under these local transformations. 
The absence of non-Abelian gauge anomaly can also be checked. $6$-dimensional anomaly polynomial for $\psi_j$ is given by ${N\over 24\pi^2}\tr(F_j^3-F_{j+1}^3)$, and thus 
they cancel completely among fermion species:
\be
\sum_{j=1}^{K}{N\over 24\pi^2}\tr(F_j^3-F_{j+1}^3)=0. 
\ee
This ensures the gauge invariance of the fermion path-integral measure. 

For $K=1$ and  $K=2$, this theory is vector-like, respectively ${\cal N}=1$ SYM and  QCD with bi-fundamental fermions, QCD(BF).
For $K\ge 3$, this theory provides a set of chiral gauge theories, and hence there is no gauge-invariant fermion mass term. 

Finally, concerning the  renormalization group  $\beta$-functions,   the one-loop  $\beta$ function coefficient for all  $K \geq 1$ are  
$\beta_0= 3N$ and the strong scale is given by  
\be
\Lambda_{1\mathrm{\mbox{-}loop}} = \mu  \rme^{-\frac{8 \pi^2 }{3g^2 N} } 
\ee 
This is the same with that of ${\cal N}=1$ SYM theory, and also with $SU(N)$ QCD with $N_f=N$-flavor fundamental quarks at the $1$-loop level.  Therefore, the one-loop definition of the strong scale in these theories are the same. 
This means that chiral quiver theories are the strongly-coupled theories, and thus it is natural to assume confinement. 
When we discuss dynamics of these theories, we assume confinement throughout this paper.

\subsection{Global chiral symmetry}
The classical symmetry of the Lagrangian (\ref{eq:quiver_lagrangian}) is given by  \footnote{Dynamics of this class of chiral gauge theories was examined first in \cite{Shifman:2008cx}. In there, global symmetry has not been fully identified correctly. As a result, despite the fact that all monopole-instanton, bion and 4d instanton amplitudes are expressed correctly and long distance effective Lagrangian on small $S^1 \times \mathbb R^3$ are correct,  the symmetries and their breaking patterns are not generally correct. Here, we identify global symmetries carefully and determine the mixed anomalies involving higher form symmetries.  As a result, all implications of the mixed anomalies agree with semi-classical analysis.} 
\be
G^{\mathrm{classical}}={U(1)^K\over (\mathbb{Z}_{N})^{K-1}}. 
\label{eq:classical_chiral}
\ee
Each $U(1)$ symmetry is defined by 
\be
\psi_j\mapsto \rme^{\im \alpha_j}\psi_j,
\ee
with $\alpha_j\sim \alpha_j+2\pi$. 
Division by $(\mathbb{Z}_N)^{K-1}$ comes from the fact that the above transformation overlaps with the center element of gauge symmetry. Indeed, $\mathbb{Z}_N\subset SU(N)_j$ acts on fermions as 
\be
\psi_j\mapsto \rme^{-{2\pi\im\over N}}\psi_j,\; \;\; \psi_{j+1} \mapsto \rme^{+{2\pi\im\over N}}\psi_{j+1}, 
\ee
and thus this is the same transformation with $\alpha_{j+1}=-\alpha_j={2\pi\over N}$. This part should be eliminated from the global symmetry. 

It is important to note that the diagonal center, 
\be
(\mathbb{Z}_N)_{\mathrm{diagonal}}\subset SU(N)_1\times SU(N)_2\times \cdots \times SU(N)_K,
\ee
does not act on the fermions at all. This is why the symmetry group is divided by $(\mathbb{Z}_N)^{K-1}$ instead of $(\mathbb{Z}_N)^K$. More importantly, as a consequence, the theory has the $1$-form symmetry,
\be
\mathbb{Z}_N^{[1]}, 
\ee
which acts on the Wilson loops, $W_j\mapsto \rme^{2\pi\im\over N}W_j$. 

Because of the Adler-Bell-Jackiw anomaly, the classical symmetry is explicitly broken to a smaller subgroup. 
We will show that the actual symmetry~\cite{Strassler:2001fs} is given by 
\be
G=\mathbb{Z}_{2N}, \quad (K=\mathrm{odd}), 
\ee
and 
\be
G=\mathbb{Z}_N\times {U(1)\over \mathbb{Z}_N}, \quad (K=\mathrm{even}). 
\ee
Accordingly, most of the $\theta$ parameters are unphysical in this theory. When $K$ is odd, all the $\theta$ angles can be eliminated by the anomalously-broken chiral transformation, so we can set $\theta_1=\cdots=\theta_K=0$. When $K$ is even, only the one combination, $\theta_1-\theta_2+\cdots-\theta_K$, is  physically meaningful, and it turns out that we can set $\theta_2=\theta_3=\cdots=\theta_K=0$ without loss of generality. 

So far, we only pay attention to the internal symmetry. We point out that these theories can have the $CP$ invariance: 
\be
\psi(x)\mapsto (\im \sigma_2) \overline{\psi}^T(x_P), \; a_\mu(x)\diff x^\mu\mapsto -a_\mu(x_P)\diff x^\mu_P, 
\ee
where $x_P:=\mathcal{P}\cdot x=(x^0,-x^1,-x^2,-x^3)$. 
The only possible origin of $CP$-violation in these models is the topological $\theta$ terms. 
Therefore, accepting the fact in the previous paragraph, theories with odd $K$ are $CP$-symmetric as we can set $\theta_i=0$ without loss of generality. 
On the other hand, theories with even $K$ have a physical $\theta$ parameter. Taking into account the $2\pi$ periodicity, they are $CP$ symmetric only at $\theta=0$ and $\theta=\pi$. 

When the gauge couplings are the same, $g_1=\cdots=g_K$, the cyclic permutation $\mathbb{Z}_K$ is also a good symmetry for odd $K$. When $K$ is even, its structure depends on the $\theta$ angle, and we shall discuss its details in Sec.~\ref{sec:permutation}.

\subsubsection{Odd $K$}

When $K$ is odd, we have the following gauge-invariant fermionic operator, 
\be
\mathcal{O}^{\mathrm{odd}}_F=\tr(\psi_1\cdots\psi_K). 
\ee
We will show that the actual symmetry is $G=\mathbb{Z}_{2N}$. The $\mathbb{Z}_{2N}$ symmetry acts on this operator as 
\be
\mathcal{O}^{\mathrm{odd}}_F\mapsto \rme^{2\pi\im\over 2N} \mathcal{O}^{\mathrm{odd}}_F, 
\ee
and its $\mathbb{Z}_2$ subgroup is identified with the fermion parity $(-1)^F$. One of the explicit realization of this symmetry generator is given by~\cite{Strassler:2001fs}
\be
\mathbb{Z}_{2N}: \psi_1\mapsto \rme^{2\pi\im \over 2N} \psi_1, \;\;\;  \psi_j\mapsto \rme^{{2\pi\im\over 2N} (-1)^j}\psi_j\,\, (j=2,\ldots,K). 
\label{eq:chiral_symmetry_oddK}
\ee
In the following of this subsection, we give a detailed derivation of this result for $K=3$. The generalization to larger odd $K$ is straightforward. 

Under the $U(1)^{K=3}$ transformation, $\psi_j\mapsto \rme^{\im \alpha_j}\psi_j$, the fermion measure $\Diff \overline{\psi}\Diff \psi$ is changed as 
\be
\Diff \overline{\psi}\Diff \psi \mapsto \Diff\overline{\psi}\Diff \psi \exp\left(\im \sum_{j=1}^{K}{N\alpha_j\over 8\pi^2}\int(\tr F_j^2+\tr F_{j+1}^2)\right). 
\label{measure-trans}
\ee
In order for this to be a good symmetry, this extra phase must be quantized in the unit of $2\pi$. 
Therefore, when $K=3$, we get the condition 
\be
N \begin{pmatrix}
1&1&0\\
0&1&1\\
1&0&1
\end{pmatrix}
\begin{pmatrix}
\alpha_1 \\ \alpha_2\\ \alpha_3
\end{pmatrix}
=
2\pi \begin{pmatrix}
k_1 \\ k_2\\ k_3
\end{pmatrix}, 
\ee
with some $k_1,k_2,k_3\in \mathbb{Z}$. We can solve this equation as 
\be
\begin{pmatrix}
\alpha_1 \\ \alpha_2\\ \alpha_3
\end{pmatrix}
=
{2\pi\over 2N}
\begin{pmatrix}
1&-1&1\\
1&1&-1\\
-1&1&1
\end{pmatrix}
\begin{pmatrix}
k_1 \\ k_2\\ k_3
\end{pmatrix}. 
\ee
At this moment, we may think that there are three independent $\mathbb{Z}_{2N}$ symmetries: $U(1)^3\to (\mathbb{Z}_{2N})^3$. Now, we must take into account the effect of divisions by gauge symmetry. 
In order to do it, it is useful to rewrite $k_1=k-k_2-k_3$ by introducing another integer $k$. We then find that 
\be
\begin{pmatrix}
\alpha_1 \\ \alpha_2\\ \alpha_3
\end{pmatrix}
={2\pi\over 2N}k \begin{pmatrix}
1\\
1\\
-1
\end{pmatrix}
+{2\pi\over N}\left\{k_2 \begin{pmatrix}
-1 \\ 0 \\ 1
\end{pmatrix}
+ k_3 \begin{pmatrix}
0 \\ -1 \\ 1
\end{pmatrix}\right\}. 
\ee
Two $\mathbb{Z}_N$ transformations given by $k_2$ and $k_3$ are nothing but the center elements of $SU(N)$ gauge symmetries, so they should be eliminated from the global symmetry. 
The $\mathbb{Z}_{2N}$ transformation given by $k$ acts faithfully on the gauge-invariant fermionic operator $\tr(\psi_1\psi_2\psi_3)$, so we find that the actual symmetry group is $G=\mathbb{Z}_{2N}$. Also note that the lowest dimensional bosonic operator charged under chiral symmetry is 
\be 
\mathcal{O}_{B}^{\mathrm{odd}}= \tr(\psi_1\cdots\psi_K\psi_1\cdots\psi_K). 
\label{eq:odd_chiral_order}
\ee
We would  expect this operator  to get a vacuum  expectation values and break the chiral symmetry down to $\mathbb{Z}_2$.

\subsubsection{Even $K$}

For even $K$, the operator 
\be
\mathcal{O}^{\mathrm{even}}=\tr(\psi_1\cdots\psi_K)
\label{poligon-even}.
\ee
is bosonic, and this is a candidate for the chiral condensate operator. In order to identify the symmetry, we also pay attention to the baryonic operators,
\be
\mathcal{B}_j\sim \psi_j^N. 
\ee

We will show that the actual symmetry is $G=\mathbb{Z}_N\times [U(1)/\mathbb{Z}_N]$. 
The $\mathbb{Z}_N$ symmetry acts as 
\be
\mathcal{O}^{\mathrm{even}}\mapsto \rme^{2\pi\im\over N}\mathcal{O}^{\mathrm{even}}, 
\ee
while the $U(1)/\mathbb{Z}_N$ symmetry acts on baryons as 
\be
\mathcal{B}_j\mapsto \rme^{\im (-1)^j \alpha}\mathcal{B}_j. 
\ee
We can realize these symmetry in the UV description as~\cite{Strassler:2001fs} 
\be
\mathbb{Z}_N: \psi_1\mapsto \rme^{2\pi\im\over N}\psi_1,\; \psi_j\mapsto \psi_j \; (j=2,\ldots,K),  
\label{eq:chiral_symmetry_evenK}
\ee
and 
\be
U(1)/\mathbb{Z}_N: \psi_j\mapsto \rme^{\im (-1)^j \alpha/N} \psi_j.  \label{eq:baryon_number_symmetry}
\ee
In the following, we explicitly show these facts when $K=4$. Again, it is straightforward to extend the proof to the case with larger even $K$. 

In order for the chiral $U(1)^{K=4}$ transformation being a good symmetry, fermion measure  \eqref{measure-trans} must remain invariant. 
Therefore,  $\alpha_j$ must satisfy 
\be
N \begin{pmatrix}
1&1&0&0\\
0&1&1&0\\
0&0&1&1\\
1&0&0&1
\end{pmatrix}
\begin{pmatrix}
\alpha_1 \\ \alpha_2\\ \alpha_3\\ \alpha_4
\end{pmatrix}
=
2\pi \begin{pmatrix}
k_1 \\ k_2\\ k_3\\ k_4
\end{pmatrix}, 
\ee
This equation can be solved if and only if $k_1 +k_3 = k_2 +k_4$, because both sides are equal to  $\sum_{j=1}^{4}\alpha_j$.  General solutions can be given using Moore-Penrose inverse, 
\be
\begin{pmatrix}
\alpha_1 \\ \alpha_2\\ \alpha_3\\ \alpha_4
\end{pmatrix}
={\alpha\over N}
\begin{pmatrix}
-1 \\ 1\\ -1\\ 1
\end{pmatrix}
+{2\pi \over 8N}
\begin{pmatrix}
3&-1&-1&3\\
3&3&-1&-1\\
-1&3&3&-1\\
-1&-1&3&3
\end{pmatrix}
\begin{pmatrix}
k_1 \\ k_2\\ k_3\\ k_4
\end{pmatrix}, 
\ee
where $\alpha\in\mathbb{R}$. Substituting $k_4=k_1-k_2+k_3$ and redefining $\alpha\to \alpha+{2\pi\over 8}(-2k_1+4k_2+2k_3)$, we find that 
\be
\begin{pmatrix}
\alpha_1 \\ \alpha_2\\ \alpha_3\\ \alpha_4
\end{pmatrix}
={\alpha\over N}
\begin{pmatrix}
-1 \\ 1\\ -1\\ 1
\end{pmatrix}
+{2\pi \over N}
\begin{pmatrix}
k_1-k_2\\
k_2\\
0\\
k_3
\end{pmatrix}. 
\ee
Here, the center gauge transformation of $SU(N)_1$ gives the identification, $k_2\sim k_2+1$, so $k_2$ can be gauged away. 
Two discrete transformations $k_1$, $k_3$ are identified by the center elements of $SU(N)_4$ gauge transformations,  i.e, $(k_1, k_3) \sim (k_1, k_3) + (1, -1)$,    so the global symmetry can be faithfully generated only by $k_1$. As a result, the actual symmetry is $(k_1,\alpha/N)\in \mathbb{Z}_N\times [U(1)/\mathbb{Z}_N]$. 
Unlike the case of odd $K$, the discrete chiral symmetry \textit{does not} include the fermion parity as its subgroup. 

We note that the $U(1)/\mathbb{Z}_N$ symmetry is a vector-like symmetry. It is easy to check that both the cubic anomaly, $U(1)^3$, and the mixed gravitational anomaly, $U(1)$-gravity-gravity, vanish. 
As a consequence, 't~Hooft anomaly matching in '80's does not apply, and does not prohibit the system to be matched by a trivial ground state.

\subsection{$\mathbb{Z}_K$ permutation symmetry}\label{sec:permutation}

When the gauge couplings are the same, $g_1=g_2=\cdots= g_K$, theories enjoy the extra $\mathbb{Z}_K$ symmetry, which cyclically permutes the $SU(N)$ gauge groups and fermion labels. 
For odd $K$, there is no physical $\theta$ angle, so this symmetry is manifest. For even $K$, the generator of $\mathbb Z_K$ acts on the physical $\theta$ angle as $\theta\rightarrow -\theta$, so the $\mathbb Z_K$ symmetry is broken explicitly by the generic $\theta$ angle to $\mathbb Z_{K/2}$, while a combination of $CP$ and $\mathbb Z_K$, $\mathbb{Z}_K^{(CP)}$, is preserved at all $\theta$ values. 
In this subsection, we shall discuss its consequences for odd and even $K$ separately. 

\subsubsection{Odd $K$}

For odd $K$, we can set $\theta_i=0$, so it is evident that we have the following $\mathbb{Z}_K$ symmetry, 
\be
a_n\mapsto a_{n+1},\; \psi_n\mapsto \psi_{n+1}, 
\ee
when $g_1=g_2=\cdots=g_K$. 
We note that this $\mathbb{Z}_K$ permutation commutes with the $\mathbb{Z}_{2N}$ discrete chiral symmetry up to gauge redundancy. 
Let us demonstrate it for $K=3$. We first apply $(\mathbb{Z}_{2N})_{\mathrm{chiral}}$ and then perform $\mathbb{Z}_{K=3}$ permutation, which gives 
\be
\begin{pmatrix}
\psi_1\\
\psi_2\\
\psi_3
\end{pmatrix} \xrightarrow{\mathbb{Z}_{2N}}
\begin{pmatrix}
\rme^{2\pi\im/N}\psi_1\\
\rme^{2\pi\im/N}\psi_2\\
\rme^{-2\pi\im/N}\psi_3
\end{pmatrix} \xrightarrow{\mathbb{Z}_{K=3}}
\begin{pmatrix}
\rme^{2\pi\im/N}\psi_2\\
\rme^{2\pi\im/N}\psi_3\\
\rme^{-2\pi\im/N}\psi_1
\end{pmatrix}. 
\ee
Applying these operations in the opposite order, we find that 
\be
\begin{pmatrix}
\psi_1\\
\psi_2\\
\psi_3
\end{pmatrix} \xrightarrow{\mathbb{Z}_{K=3}}
\begin{pmatrix}
\psi_2\\
\psi_3\\
\psi_1
\end{pmatrix} \xrightarrow{\mathbb{Z}_{2N}}
\begin{pmatrix}
\rme^{2\pi\im/N}\psi_2\\
\rme^{-2\pi\im/N}\psi_3\\
\rme^{2\pi\im/N}\psi_1
\end{pmatrix} 
\sim \begin{pmatrix}
\rme^{2\pi\im/N}\psi_2\\
\rme^{2\pi\im/N}\psi_3\\
\rme^{-2\pi\im/N}\psi_1
\end{pmatrix}. 
\ee
At the last step, we use the identification by the center gauge transformation $\mathbb{Z}_N\subset SU(N)_1$, so these two symmetry transformations commute on gauge-invariant local operators. 
This permutation symmetry does not have an anomaly, so it is consistent to assume that the ground states respect the $\mathbb{Z}_K$ symmetry\footnote{An example for the order parameter of $\mathbb{Z}_K$ is $\sum_{\ell} \rme^{{2\pi\im\over K}\ell} \tr(F_{\ell,\mu\nu}^2)$. This operator is singlet under other symmetries, $CP$ and $\mathbb{Z}_{2N}$ chiral, so it is a good order parameter for spontaneous breakdown of $\mathbb{Z}_K$. }. 

Especially, we would like to note that the chiral order parameter (\ref{eq:odd_chiral_order}) can be made $\mathbb{Z}_K$-singlet. 
In order to show it, we must be careful about the spinor indices, so we introduce the following notation: 
\be
[\xi \eta]\equiv \ve_{\alpha\beta} \xi_\alpha \eta_\beta, 
\ee
where $\xi_\alpha, \eta_\beta$ are undotted spinors. Using this, we take the following contraction of spinor indices for the chiral order parameter\footnote{There are many other possibilities for spinor contractions, but we here take the simplest one in terms of the notational issue. }, 
\be
\mathcal{O}_B^{\mathrm{odd}}=\tr([\psi_1\psi_2] \cdots [\psi_K \psi_1] \cdots [\psi_{K-1}\psi_K]). 
\ee
This is an example of $\mathbb{Z}_K$-singlet chiral order parameter: 
\be
\mathcal{O}_B^{\mathrm{odd}} \xrightarrow{\mathbb{Z}_K} \tr([\psi_2\psi_3] \cdots [\psi_1 \psi_2] \cdots [\psi_{K}\psi_1]) = \tr([\psi_1\psi_2] \cdots [\psi_K \psi_1] \cdots [\psi_{K-1}\psi_K])=\mathcal{O}_B^{\mathrm{odd}}. 
\ee
Here, we use the cyclic property of the trace operation. As a result, $\mathcal{O}_B^{\mathrm{odd}}$ is a good order parameter of chiral symmetry breaking, since it can develop the nonzero expectation value with unbroken $\mathbb{Z}_K$ symmetry. 

\subsubsection{Even $K$}

For a moment, let us assume the case $\theta=0$. In this case, the analysis becomes quite similar to that of the odd-$K$ case. We can readily find that $\mathbb{Z}_K$ permutation symmetry commutes with the $\mathbb{Z}_N$ discrete chiral symmetry but has the structure of semidirect product with $U(1)/\mathbb{Z}_N$. 
At $\theta=0$, this symmetry does not enter the 't~Hooft anomaly, so we will assume that $\mathbb{Z}_K$ is unbroken. 

To find the consequence of unbroken $\mathbb{Z}_K$ symmetry, we again need to specify the spinor contractions of chiral operators (\ref{poligon-even}). Interestingly, for the case of even $K$, we can have two different contractions even if we restrict ourselves to contractions of neighboring spinors in (\ref{poligon-even}):
\bea
\mathcal{O}_{(1)}^{\mathrm{even}}&=&\tr([\psi_1\psi_2]\cdots [\psi_{K-1}\psi_K]), \\
\mathcal{O}_{(2)}^{\mathrm{even}}&=&\tr([\psi_2\psi_3]\cdots [\psi_{K}\psi_1]). 
\eea
Under the $\mathbb{Z}_K$ permutation, these two operators are exchanged,
\be
\mathcal{O}_{(1)}^{\mathrm{even}} \xleftrightarrow{\mathbb{Z}_K} \mathcal{O}_{(2)}^{\mathrm{even}}. 
\ee
Therefore, under the assumption of unbroken $\mathbb{Z}_K$, these two operators should have the same expectation value at $\theta=0$:
\be
\left.\langle \mathcal{O}_{(1)}^{\mathrm{even}}\rangle\right|_{\theta=0} =\left.\langle \mathcal{O}_{(2)}^{\mathrm{even}} \rangle\right|_{\theta=0}. 
\ee

From now on, let us turn on the $\theta$ angle. In our convention, we take $\theta_1=\theta$, and $\theta_{i\not=1}=0$. When we apply the $\mathbb{Z}_K$ permutation, this is mapped to $\theta_2=\theta$ and $\theta_{i\not=2}=0$. 
We can change the location of this non-zero $\theta$ angle to $\theta_1$ by using the anomalous $U(1)$ chiral rotation on $\psi_1$, and we get $\theta_1=-\theta$ and $\theta_{i\not = 1}=0$. 
As a result, as in the case of $CP$ transformation, the physical $\theta$ angle flips its sign under $\mathbb{Z}_K$, so it is a good symmetry only at $\theta=0$ or $\theta=\pi$. 

This argument also shows that the combination of $\mathbb{Z}_K$ permutation and $CP$ transformation is a good symmetry at any values of $\theta$. We denote this operation as $\mathbb{Z}_K^{(CP)}$, and it is consistent to assume that $\mathbb{Z}_K^{(CP)}$ is unbroken at any $\theta$ angle. 
As the $CP$ transformation do not commute with the discrete chiral symmetry, the rigorous statement is that, when the discrete chiral symmetry is spontaneously broken, each vacuum respects $\mathbb{Z}_K^{(CP)}$ after appropriate rotation of broken chiral symmetry.

\section{Vacuum structure of odd-sites quiver theories}\label{sec:odd_sites}

In this section, we study the possible low-energy behavior of chiral quiver theory when $K$ is odd. 
We first derive the mixed 't~Hooft anomaly between the center symmetry, $\mathbb{Z}_N^{[1]}$, and the chiral symmetry $\mathbb{Z}_{2N}$. As a consequence of anomaly matching condition, chiral symmetry must be spontaneously broken when Wilson loops are confined. 
Considering the double-trance deformation on $\mathbb{R}^3\times S^1$, we confirm this prediction by reliable semiclassics, and we find $N$ distinct vacua by spontaneous chiral symmetry breaking. 

\subsection{'t~Hooft anomaly matching}\label{sec:odd_anomaly}

Let us show that there is an 't~Hooft anomaly between the center and discrete chiral symmetries: 
\be
(\mathbb{Z}_N^{[1]})_{\mathrm{center}}\times (\mathbb{Z}_{2N})_{\mathrm{chiral}}. 
\ee
In order to see it, we introduce the background $\mathbb{Z}_N$ two-form gauge field $B$. Following the convention in Ref.~\cite{Kapustin:2014gua}, we realize such a discrete $2$-form gauge field by a pair of $U(1)$ $2$-form and $1$-form gauge fields $(B,C)$ satisfying
\be
NB=\diff C. 
\ee
We embed the $SU(N)$ gauge fields $a_j$ into $U(N)$ gauge fields $\widetilde{a}_j$, which locally satisfies 
\be
\widetilde{a}_j=a_j+{1\over N}C \bm{1}. 
\ee
In order for the structure group being $[SU(N)_1\times \cdots \times SU(N)_K]/(\mathbb{Z}_N)_{\mathrm{diag}}$, we need to postulate the invariance under the $1$-form gauge transformation, 
\be
B\mapsto B+\diff \lambda,\; \widetilde{a}_j\mapsto \widetilde{a}_j + \lambda \bm{1}. 
\ee
We note that the covariant derivative,
\be
D\psi_j=\diff \psi_j+\im\, \widetilde{a}_j \psi_j-\im\, \psi_j \widetilde{a}_{j+1}, 
\ee
manifestly satisfies the invariance under the $1$-form transformation. In order to maintain the invariance of the gauge kinetic term, we replace the field strength by 
\be
\widetilde{F}_j-B\bm{1}:=(\diff \widetilde{a}_j+\im\, \widetilde{a}_j\wedge \widetilde{a}_j)-B\bm{1}. 
\ee
With this setup, we can compute the partition function by integrating over fermonic fields and gauge fields, and we denote it by $\mathcal{Z}[B]$. 

In order to see the 't~Hooft anomaly, we perform the discrete chiral transformation (\ref{eq:chiral_symmetry_oddK}) on $\mathcal{Z}[B]$. 
This transformation is given by $\alpha_1={2\pi\over 2N}$ and $\alpha_j={2\pi\over 2N}(-1)^j$ for $j=2,\ldots, K$, 
so the fermion measure gets the phase
\bea
&&{2\pi\im\over 2N}{N \over 8\pi^2}\int \left\{(\tr(\widetilde{F}_1-B)^2+\tr(\widetilde{F}_2-B)^2)+\sum_{j=2}^K (-1)^j(\tr(\widetilde{F}_j-B)^2+\tr(\widetilde{F}_{j+1}-B)^2)\right\}\nonumber\\
&=&{\im\over 4\pi}\int \tr (\widetilde{F}_2-B)^2. 
\eea
Using the $U(N)$ index theorem on spin manifolds, we find that the partition function $\mathcal{Z}[B]$ acquires the overall phase under the chiral symmetry as 
\be
(\mathbb{Z}_{2N})_{\mathrm{chiral}}: \mathcal{Z}[B]\mapsto \exp\left(-{\im N\over 4\pi}\int B\wedge B\right)\mathcal{Z}[B]. 
\label{eq:anomaly_oddK}
\ee

As a consequence, when we  also introduce the discrete chiral gauge field $A_\chi^{\mathrm{odd}}$, satisfying 
\be
2N A_\chi^{\mathrm{odd}}=\diff \phi, 
\ee 
with $2\pi$-periodic scalar $\phi$, the partition function has the gauge ambiguity canceled by anomaly inflow. The corresponding $5$-dimensional topological action is 
\be
S_{5\mathrm{d}}={2\pi \im\over N} \int {2N\over 2\pi} A_\chi^{\mathrm{odd}}\wedge {N^2\over 8\pi^2} (B\wedge B). 
\ee
This topological action defines a $5$-dimensional symmetry-protected topological (SPT) phase of the class $\mathbb{Z}_N$, so that the $\mathbb Z_2$ part of $\mathbb Z_{2N}$ chiral symmetry is not anomalous.
In order to match the anomaly, we need to require a nontrivial low-energy dynamics, such as 
\begin{itemize}
\item massless excitations, e.g. in Coulomb phase, 
\item $N$ vacua by discrete chiral symmetry breaking, $(\mathbb{Z}_{2N})_{\mathrm{chiral}}\to \mathbb{Z}_2$, or
\item $\mathbb{Z}_N$ topological order by deconfinement, $(\mathbb{Z}_N^{[1]})_{\mathrm{center}}\to 1$. 
\end{itemize}
Since the beta function of this gauge theory is the same with that of $\mathcal{N}=1$ $SU(N)$ super Yang-Mills theory, we can expect confinement with non-zero mass gap. 
Under this reasonable assumption, anomaly matching requires the chiral symmetry breaking,
\be
(\mathbb{Z}_{2N})_{\mathrm{chiral}}\to \mathbb{Z}_2. 
\ee
We emphasize that both symmetry and anomaly are the same with those of $\mathcal{N}=1$ $SU(N)$ SYM theory given in Refs.~\cite{Komargodski:2017smk, Shimizu:2017asf}. 
Although the nonperturbative orbifold conjecture~\cite{Schmaltz:1998bg, Strassler:2001fs} itself does not work~\cite{Kovtun:2004bz, Kovtun:2005kh, Kovtun:2007py}, these theories share the surprisingly similar structure. Indeed, as we shall see in Sec.~\ref{sec:universality}, all odd site quiver theories can be deformed to SYM. 

\subsection{Semiclassics on small $\mathbb{R}^3\times S^1$}
We consider the double-trace deformation of the $[SU(N)]^K$ chiral quiver theory, so that the zero-form part of the center-symmetry acting on 
gauge holonomy around the $S^1$ circle is unbroken.   
Most part of this section has been done in Ref.~\cite{Shifman:2008cx}, but the global nature of theories is not correctly captured there. We carefully perform their analysis again in order to give correct identifications of vacuum structures.

Because of the double-trace deformation, the gauge invariance reduces to the maximal abelian subgroup and Weyl permutations, $[U(1)^{N-1} \rtimes  W_{\mathfrak{su}(N)} ]^K $. 
The double-trace term forces the center-symmetric vacuum, so that the Polyakov-loop phase $\bm{\phi}$ takes the vacuum-expectation value, $\bm{\phi}\sim {2\pi\over N}(1,2,\ldots, N)$, and this fixes the gauge for $W_{\mathfrak{su}(N)}$. 
The dual photon part of the action (ignoring the non-perturbative effects) is given by\footnote{ In order to introduce dual photons by $3$d Abelian duality to Cartan part of gluons, all the electrically charged excitations under $U(1)^{N-1}$ must be gapped. In the chiral quiver theory,  the diagonal components of bifundamental fermions do not acquire mass from the center-symmetric Polyakov loop, so we need to introduce the real mass by taking the twisted boundary condition using (\ref{eq:chiral_symmetry_oddK}) and \eqref{eq:baryon_number_symmetry}, for $K$ odd and even, respectively.  } 
\be
S= \sum_{j=1}^{K} {1\over L}\int_{M_3}\left( {g^2\over 16\pi^2}\left|\diff \bm{\sigma}_{j}\right|^2\right). 
\ee
where  $\bm{\sigma}_{j}$ is the dual photon field associated with the $j^{\rm th}$ gauge group factor.  The $\bm{\sigma}_{j}$   field lives in  
\be
\bm{\sigma}_j \in {\mathbb{R}^{N-1}\over  (2\pi \Lambda_w) }, 
\ee
where $\Lambda_w$ is the weight lattice, 
but further gauge identification for $(\bm{\sigma}_{1}, \ldots, \bm{\sigma}_{K})$-tuple  is described below.

The monopole-instanton operators associated with the $j$-th gauge group factor are given by 
\be
{\cal M}_{j,i}(x)=\rme^{-S_I/N} \rme^{\im \bm{\alpha}_i\cdot \bm{\sigma}_{j}(x)}    (\bm{\alpha}_i\cdot {\bm \psi}_{j-1}) (\bm{\alpha}_i\cdot {\bm \psi}_{j}),     \quad (i=1,\ldots, N, \;\;  j=1, \ldots K) 
\label{monopole}
\ee
Here, $S_I={8\pi^2/g^2}$ is the instanton action.  The fermion zero mode structure is dictated by the index theorem for the Dirac operator on $\mathbb R^3 \times S^1$ \cite{Nye:2000eg,Poppitz:2008hr}.  The zero modes are  associated with the incoming/outgoing  fermion link field on the quiver. 

It is worth noting that the monopole operator is invariant under a gauge transformation associated with the $j$-th gauge group factor, but not with
$(j-1)$-th and $(j+1)$-th gauge groups.  As a consequence, these events themselves cannot contribute to the effective Lagrangian, and this point is an important difference from semiclassics for $\mathcal{N}=1$ SYM on $\mathbb{R}^3\times S^1$. However, they are useful as building block of gauge-invariant nonperturbative contributions to the dynamics. 
   
 At second order in semi-classics, we have magnetic bion contributions, corresponding to the correlated events $ {\cal B}_{j,i}=  [{\cal M}_{j,i}(x) \overline  {\cal M}_{j,i+1}(x)]$ where the zero modes of the monopole  are soaked up by the zero modes of the  anti-monopole. The dilute gas of the magnetic bions induce a potential: 
\bea
V^{(2)}&=&-\rme^{-2S_I/N} \sum_{i=1}^{N} \sum_{j=1}^{K}  \cos \left( (\bm{\alpha}_i -\bm{\alpha}_{i+1}) \cdot \bm{\sigma}_{j}(x)   \right) 
\label{bion-pot}
\eea
The minima of the potential are located at  
\bea 
(\bm{\sigma}_{1}, \ldots, \bm{\sigma}_{K}) = \frac{2 \pi }{N} {\bm \rho} \;  (k_1, \ldots, k_K)  \qquad  k_j= 1, \ldots, N
\label{minima}
\eea 
where $ {\bm \rho} = \sum_{n=1}^{N-1} {\bm \mu}_n $ is the Weyl vector.  There are $N^K$ minima of this potential,  but below we show that they fall into $N$ gauge inequivalent orbits, hence there are only $N$ vacua in the quantum theory. 
The $  (\mathbb{Z}_{N})^{K-1} $ part of the gauge redundancy provides the following extra identifications on the 
 $(\bm{\sigma}_{1}, \ldots, \bm{\sigma}_{K})$-tuple.  
The  $  (\mathbb{Z}_{N})^{K-1} $ gauge identifications are 
\bea 
(\bm{\sigma}_{1}, \ldots, \bm{\sigma}_{K}) \sim (\bm{\sigma}_{1}, \ldots, \bm{\sigma}_{K}) +  \frac{2 \pi }{N} {\bm \rho} \Big[ (1, -1, \ldots, 0) l_1+ 
\ldots + (0,  \ldots, 0, 1,-1 ) l_{K-1} \Big], \qquad \qquad 
\label{gaugeiden}
\eea 
with $l_i\in \mathbb{Z}$. 
After these identifications,  we observe that there are $N$ true vacua in the $[SU(N)]^K$ theory 
(and in the  gauge orbit of each true vacuum, there are $N^{K-1}$ gauge equivalent copies). 
In order to take into account these facts, it is useful to consider the gauge-invariant combination
 \bea 
\sum_{j=1}^{K} \bm{\sigma}_{j}  =  \frac{2 \pi }{N} {\bm \rho} k, \qquad k \equiv \Big( \sum_{j=1}^{K} k_j  \Big) \bmod N . 
\label{gauge-invariant}
\eea

 A quicker way to reach to the same conclusion is as follows. All monopole events ${\cal M}_{j,i}(x)$ are gauge covariant but not invariant, so their product 
 $ \prod_{j=1}^{K} {\cal M}_{j,i}(x)$  is the leading gauge-invariant nonperturbative event with fermionic zero modes.  This combination is  given by 
  \bea 
  {\cal {MR}}_i \equiv 
\prod_{j=1}^{K} {\cal M}_{j,i} \sim  \rme^{- K S_I/N}   \rme^{\im  \bm{\alpha}_i\cdot  \left( \sum_{j=1}^{K}  \bm{\sigma}_{j}(x) \right)}   \left( ({\alpha}_i\cdot \psi_1) \cdots  ({\alpha}_i\cdot \psi_K)    \right)^2
\label{double-polygon}
\eea 
The $\mathbb Z_{2N}$  chiral transformation  \eqref{eq:chiral_symmetry_oddK}  acts on the multi-fermion operators 
as   $\tr \left( \psi_1 \ldots  \psi_K \right)^2 \rightarrow \rme^{2\pi\im \over N} \tr \left( \psi_1 \ldots  \psi_K \right)^2 $. 
Since $\mathbb Z_{2N}$  is a genuine symmetry of the theory, it must be respected by the monopole operators. 
This demands that under a chiral rotation,   the pure  flux part of the monopole operator transform by a $\mathbb Z_{N}$ phase. 
\bea
\rme^{\im  \bm{\alpha}_i\cdot  \left( \sum_{j=1}^{K}  \bm{\sigma}_{j}(x) \right)}  \rightarrow   \rme^{-{2\pi\im \over N}}  \rme^{\im  \bm{\alpha}_i\cdot  \left( \sum_{j=1}^{K}  \bm{\sigma}_{j}(x) \right)} 
 \eea 
 The exponent is noting but the gauge invariant combination \eqref{gauge-invariant}. The bion induced potential provides a vacuum expectation value for the chiral order parameter, giving the $N$-vacua: 
 \bea
\left \langle \rme^{\im  \bm{\alpha}_i\cdot  \left( \sum_{j=1}^{K}  \bm{\sigma}_{j}(x) \right)}   \right \rangle  = \rme^{{2\pi\im  k \over N}}, \qquad k=1, \ldots, N.
\label{condensate-smallL}
 \eea

 On $\mathbb R^4$,  a natural order parameter for the chiral symmetry breaking is the double-polygon operator, which is a bosonic gauge invariant operator with $2K$ fermion insertions: 
 \be
\mathcal{O}^{\mathrm{odd}}_{B}=\tr([\psi_1\psi_2]\cdots[\psi_K \psi_1]\cdots [\psi_{K-1}\psi_K] ).
\ee  
Under the chiral transformation \eqref{eq:chiral_symmetry_oddK}, this operator transform as  $ \mathcal{O}_B^{\mathrm{odd}} \rightarrow   \rme^{{2\pi\im  \over N}}  \mathcal{O}_B^{\mathrm{odd}}$ and its condensation lead to the spontaneous discrete chiral symmetry breaking  $\mathbb{Z}_{2N} \rightarrow  \mathbb{Z}_{2} $. 
On small $\mathbb R^3 \times S^1$, the fermion zero mode structure of the  double-polygon operator \eqref{double-polygon} is exactly right 
to contribute to this condensate.  We showed that in the  vacuum, it is the flux part of the double-polygon operator   \eqref{condensate-smallL} that acquires an expectation value and breaks chiral symmetry. In this $k$-th vacuum, call it $| \Omega_k \rangle$, 
we can calculate the expectation value of $\mathcal{O}^{\mathrm{odd}}_{B}$ as well. 
We find that   $ \prod_{j=1}^{K} {\cal M}_{j,i}(x)$  for all $i=1, \ldots, N$ contributes to it. We find, up to inessential numerical factors, 
\bea 
 \langle  \Omega_k| \mathcal{O}^{\mathrm{odd}}_{B} | \Omega_k \rangle = N L^{-3K} \rme^{ -  K \frac{8 \pi^2}{g^2 N}} \rme^{{2\pi\im  k \over N}}. 
\eea
The $\beta$ function of the chiral quiver theory is the same as ${\cal N}=1$ SYM in the large-$N$ limit. 
Using the one-loop expression, with the renormalization scale $\mu\sim L^{-1}$, we find that  
\bea 
 \langle  \Omega_k| \mathcal{O}^{\mathrm{odd}}_{B} | \Omega_k \rangle = N \Lambda^{3K} \rme^{{2\pi\im  k \over N}}, 
\eea
Despite the fact that our analysis is in the weakly coupled domain, it produces a chiral condensate which is dictated by the strong scale $\Lambda$ of the theory.  At the leading order, just like ${\cal N}=1$ SYM theory, the $L$ dependence of the chiral condensate disappears.

\section{Vacuum structure of even-sites quiver theories}\label{sec:even_sites}

In this section, we study the low-energy dynamics for the case $K$ is even. 
We first derive mixed 't~Hooft anomaly between $\mathbb{Z}_{N}^{[1]}$ and $\mathbb{Z}_N\times [U(1)/\mathbb{Z}_N]$, and hence anomaly matching suggests the $N$ isolated vacua by chiral symmetry breaking when assuming confinement. 
We next derive the generalized anomaly, or global inconsistency, related to the $\theta$ angle periodicity. As a consequence, not only are there $N$ vacua due to the chiral symmetry breaking, the system also has an $N$-branch structure, and those branches are exchanged as we dial the $\theta$ angle. 
Both of these observations are confirmed by reliable semiclassics on $\mathbb{R}^3\times S^1$. 

\subsection{'t~Hooft anomaly matching for chiral symmetry}\label{sec:anomaly_even_sites}

We introduce the background gauge field for the vector-like symmetry,
\be
\mathbb{Z}_{N}^{[1]}\times {U(1)\over \mathbb{Z}_N}. 
\ee
The contents of the background gauge field are given by 
\begin{itemize}
\item $\mathbb{Z}_N$ $2$-form gauge field, $(B,C)$ with $NB=\diff C$, for $\mathbb{Z}_{N}^{[1]}$, 
\item $U(1)$ gauge field $A$,
\item $\mathbb{Z}_N$ $2$-form gauge field, $(B',C')$ with $NB'=\diff C'$, to take the quotient $U(1)/\mathbb{Z}_N$. 
\end{itemize}
Dynamical $SU(N)$ gauge fields $a_j$ are embedded into $U(N)$ gauge fields, locally written as 
\be
\widetilde{a}_{2n-1}=a_{2n-1}+{1\over N}C\bm{1},\;\; \widetilde{a}_{2n}=a_{2n}+{1\over N}C\bm{1}+{1\over N}C'\bm{1},
\ee
for $n=1,\ldots,K/2$. 
We postulate the invariance under $1$-form gauge transformations, 
\be
B\mapsto B+\diff \lambda,\;\; B'\mapsto B'+\diff \lambda',
\ee
and 
\be
\widetilde{a}_{2n-1}\mapsto \widetilde{a}_{2n-1}+\lambda \bm{1},\;\; \widetilde{a}_{2n}\mapsto \widetilde{a}_{2n}+\lambda\bm{1}+\lambda'\bm{1},\;\; A\mapsto A-\lambda'. 
\ee
Since the covariant derivatives on fermionic fields are now given as 
\be
D\psi_{j}=\diff \psi_{j}+\im\, \widetilde{a}_{j}\psi_{j}-\im\, \psi_{j} \widetilde{a}_{j+1}+\im (-1)^j A \psi_{j},
\ee
the manifest $1$-form gauge invariance holds. 
$SU(N)$ field strengths should be replaced by the $U(N)$ field strengths as 
\be
F_{2n-1}\to \widetilde{F}_{2n-1}-B\bm{1},\;\; F_{2n}\to \widetilde{F}_{2n}-B\bm{1}-B' \bm{1}. 
\ee
We also note that $N(\diff A+B')$ is the field strength for $U(1)/\mathbb{Z}_N$. This can be identified as the gauge field for the baryon number symmetry in the correct canonical normalization, and we denote that 
\be
\diff A_\mathrm{B}:= N (\diff A + B'). 
\ee
This identification of the baryon-number gauge field has been also used in QCD with fundamental quarks, and see Ref.~\cite{Tanizaki:2018wtg}. 

Now, in order to find the mixed 't~Hooft anomaly, we perform the $\mathbb{Z}_N$ transformation, $\psi_1\mapsto \rme^{2\pi\im/N}\psi_1$. 
Under this transformation, the fermion measure acquires the phase factor~\cite{Tanizaki:2018wtg}, 
\bea
&&{2\pi\im \over N}{1\over 8\pi^2}\int \tr_{1,2}\left(\widetilde{F}_1-\widetilde{F}_2^T-\diff A\right)^2\nonumber\\
&=&{2\pi\im \over N}{1\over 8\pi^2}\int \left(N \tr (\widetilde{F}_1-B)^2+N \tr(\widetilde{F}_{2}-B-B')^2+N^2 (\diff A +B')^2\right)\nonumber\\
&=&-\im\left({N \over 2\pi}\int B\wedge B+{1\over 2\pi}\int B\wedge \diff A_\mathrm{B}\right),  
\eea
mod $2\pi\im$. Therefore, the partition function $\mathcal{Z}[B,A_B]$ under the background gauge fields is transformed as 
\be
(\mathbb{Z}_N)_{\mathrm{chiral}}:\mathcal{Z}[B,A_\mathrm{B}]\mapsto \exp\left(-\im{1 \over 2\pi}\int (N B\wedge B+B\wedge \diff A_\mathrm{B})\right)\mathcal{Z}[B,A_\mathrm{B}]. 
\label{eq:anomaly_evenK}
\ee

This means that the partition function suffers from gauge ambiguity when we also introduce the background chiral gauge field $A_\chi^{\mathrm{even}}$. 
Since the discrete chiral symmetry for even $K$ is $\mathbb{Z}_N$, it satisfies 
\be
N A_\chi^{\mathrm{even}}=\diff \phi, 
\ee
with some $2\pi$-periodic scalar field $\phi$. The $5$-dimensional SPT action is given by 
\be
S_{5\mathrm{d}}={4\pi\im\over N}\int {N\over 2\pi}A_\chi^{\mathrm{even}}\wedge {N^2\over 8\pi^2} (B\wedge B) + {2\pi\im\over N}\int {N\over 2\pi} A_\chi^{\mathrm{even}}\wedge {N\over 2\pi}B\wedge {1\over 2\pi} \diff A_\mathrm{B}. 
\label{eq:even_K_5d_anomaly}
\ee
Let us discuss how the low-energy physics can match this anomaly. 

Since the $\beta$ function of the gauge coupling is again the same with that of $\mathcal{N}=1$ super Yang-Mills theory, it is natural to assume confinement. 
The first term of (\ref{eq:even_K_5d_anomaly}) only involves the center and discrete chiral symmetries, and the second one involves $U(1)/\mathbb{Z}_N$ as well. 
When $N$ is odd, the center-chiral mixed anomaly requires the complete chiral symmetry breaking, $(\mathbb{Z}_N)_{\mathrm{chiral}}\to 1$, and this matches the second one automatically. 
Therefore, under the assumption of confinement there must be $N$ isolated vacua by the chiral symmetry breaking. 

When $N$ is even, the story is slightly more complicated. 
The center-chiral mixed anomaly only requires $N/2$ vacua by chiral symmetry breaking $(\mathbb{Z}_N)_{\mathrm{chiral}}\to \mathbb{Z}_2$, because this $\mathbb{Z}_2$ subgroup is neutral under the first term of (\ref{eq:even_K_5d_anomaly})\footnote{When we put the theory on non-spin manifolds, such as $\mathbb{C}P^2$, by introducing a kind of spin-charge relation, we can obtain stronger constraints from center-chiral mixed anomaly~\cite{Anber:2020gig}. 
The index of Dirac operators on $\mathbb{C}P^2$ with minimal non-trivial 't~Hooft flux turns out to be $(1+N/2)$ mod $N$, if we do not introduce any flux on $U(1)/\mathbb{Z}_N$ for even $N$. 
Therefore, when $N\in 4\mathbb{Z}$, the stronger constraint indeed appears, and we can claim complete chiral symmetry breaking just from center-chiral mixed anomaly. 
For $N\in 4\mathbb{Z}+2$, the center-chiral mixed anomaly is still order $N/2$.

We thank Mohamed Anber for pointing out the usefulness of non-spin manifolds, with detailed explanations on how such computations can be done.
}. 
In order to match the second term, anomaly matching requires either full chiral symmetry breaking $(\mathbb{Z}_N)_{\mathrm{chiral}}\to 1$ or some gappless excitations charged under $U(1)/\mathbb{Z}_N$ with partial chiral symmetry breaking $(\mathbb{Z}_N)_{\mathrm{chiral}}\to \mathbb{Z}_2$.  
When $N$ is not too small, we believe that the low-energy behavior is controlled by smooth large-$N$ limit, and the different behaviors between even and odd $N$ is difficult to imagine. 
Therefore, we conclude that anomaly matching is satisfied by $N$ isolated vacua by the spontaneous chiral symmetry breaking,  
\be
(\mathbb{Z}_N)_{\mathrm{chiral}}\to 1
\ee 
for both even and odd $N$. 

We note, however, that the theory with $N=2$ may be special. In this case, the first term of (\ref{eq:even_K_5d_anomaly}) is completely trivial, and thus the chiral symmetry breaking does not have to occur even partially. 
To match the second term, the existence of gappless excitations charged under $U(1)/\mathbb{Z}_2$ should be sufficient. 

\subsection{Generalized 't~Hooft anomaly for $\theta$ periodicity}

For even-sites quiver theories, we have one physical $\theta$ parameter. Without loss of generality, we can take $\theta_1=\theta$ and $\theta_2=\cdots=\theta_K=0$, where $\theta \sim \theta+2\pi$. 

Assuming confinement, we can find an interesting phase structure as we dial $\theta$. In order to see it, we discuss the global inconsistency~\cite{Gaiotto:2017yup, Tanizaki:2017bam, Kikuchi:2017pcp}, or generalized 't~Hooft anomaly~\cite{Cordova:2019jnf,Cordova:2019uob}, for the periodicity of $\theta$. 
Let us again introduce the background gauge field for the vector-like symmetry, $\mathbb{Z}_N^{[1]}\times [U(1)/\mathbb{Z}_N]$, and then we compare the partition function at $\theta$ and $\theta+2\pi$. 
Since the topological charge for the $SU(N)_1$ gauge group is given by 
\be
{1\over 8\pi^2}\int \tr(\widetilde{F}_1-B)^2={1\over 8\pi^2}\int \tr(\widetilde{F}_1^2)-{N\over 8\pi^2}\int B^2. 
\ee
The first term is quantized to integers due to the $U(N)$ index theorem on spin manifolds. As a consequence, we find that 
\be
\mathcal{Z}_{\theta+2\pi}[B,A_\mathrm{B}]= \exp\left(\im {N\over 4\pi}\int B\wedge B\right)\mathcal{Z}_{\theta}[B,A_\mathrm{B}]. 
\label{eq:generalized_anomaly_theta}
\ee
This leads to the $N$ branch structure, and phase transitions must happen at least once while we change $\theta$ continuously from $0$ to $2\pi$ . 

A natural location of this first-order phase transition would be $\theta=\pi$. We can justify this expectation by paying attention to the $CP$ symmetry. 
In this model, the only origin of $CP$ violation is the $\theta$ angle, as other couplings are manifestly real. 
Since $CP$ transformation effectively flips the sign of $\theta$, the $CP$ symmetry exists only at $\theta=0$ and $\theta=\pi$ thanks to the $2\pi$ periodicity. 

By using the relation (\ref{eq:generalized_anomaly_theta}), we can find the mixed anomaly, or global inconsistency, for the $\mathbb{Z}_{N}^{[1]}$ symmetry and the $CP$ symmetry at $\theta=\pi$. Indeed, we obtain that 
\be
\mathcal{Z}_{\theta=\pi}[B,A_{\mathrm{B}}]\mapsto  \exp\left(-\im {N\over 4\pi}\int B\wedge B\right) \mathcal{Z}_{\theta=\pi}[B,A_{\mathrm{B}}]. 
\ee
When $N$ is even, there is no local counter term that can cancel this anomalous phase, so we get the genuine anomaly. As a consequence, we conclude the spontaneous $CP$ breaking at $\theta=\pi$ for even $N$. 
Therefore, there are $2N$ vacua at $\theta=\pi$, and the symmetry breaking pattern is 
\be
(\mathbb{Z}_N)_{\mathrm{chiral}}\rtimes (\mathbb{Z}_2)_{CP}\to 1. 
\ee

When $N$ is odd, there is a local counter term that can cancel this anomaly, so we must carefully compare the gauged partition functions at $\theta=0$ and $\theta=\pi$. 
Basically, there are two options for low-energy dynamics in this situations under the assumption of confinement~\cite{Gaiotto:2017yup, Tanizaki:2017bam, Kikuchi:2017pcp}:
\begin{itemize}
\item $CP$ is spontaneously broken either at $\theta=0$ or $\theta=\pi$, or 
\item phases at $\theta=0,\,\pi$ are distinct as $\mathbb{Z}_N^{[1]}$-protected SPT phases. 
\end{itemize}
However, we again expect the smooth large-$N$ behavior, so it is natural to expect that $CP$ is broken at $\theta=\pi$ also for odd $N$.

When $g_1=\cdots=g_K$, we also note that this conclusion is consistent with the mixed anomaly involving $\mathbb{Z}_K$ permutation at $\theta=\pi$. As we discussed in Sec.~\ref{sec:permutation}, the full $\mathbb{Z}_K$ permutation is a good symmetry only at $\theta=0$ or $\theta=\pi$, and the transformation at $\theta=\pi$ is given by the permutation, $\psi_n\mapsto \psi_{n+1}$, followed by the chiral transformation, $\psi_1\mapsto \rme^{-2\pi \im/N}\psi_1$. Since this effectively flips the sign of $\theta=\pi $, we obtain the mixed anomaly,
\be
(\mathbb{Z}_K)_{\mathrm{permutation}}: \mathcal{Z}_{\theta=\pi}[B,A_\mathrm{B}]\mapsto \exp\left({\im N\over 4\pi}\int B\wedge B\right)\mathcal{Z}[B,A_\mathrm{B}]. 
\ee
We note that this anomaly has the same anomalous phase with that of mixed $CP$ anomaly, which means that $\mathbb{Z}_K^{(CP)}$ does not have an anomaly. Another important point is that $\mathbb{Z}_K$ does not act on $B$, so there is no possible local counter term for this $\mathbb{Z}_K$ mixed anomaly, unlike the case of $CP$ symmetry. 
Therefore it is quite natural to assume that these anomalies are matched by spontaneous breakdown of $\mathbb{Z}_K\times CP$ down to $\mathbb{Z}_K^{(CP)}$. 

For even $K$, the anomaly structure, including the symmetry group, is the same with that of QCD(BF) at the massless point (see, e.g., Refs.~\cite{Tanizaki:2017bam, Karasik:2019bxn, Wan:2019oax} for QCD(BF), and also Refs.~\cite{Tanizaki:2017mtm, Tanizaki:2018wtg} for related anomaly in QCD with fundamental quarks). 
Again, this class of chiral gauge theories shows very similar features with the vector-like theory, i.e. QCD(BF) (see also Sec.~\ref{sec:universality}). 

\subsection{Semiclassics on small $\mathbb{R}^3\times S^1$}
The semiclassics with double-trace deformation has been analyzed in Ref.~\cite{Shifman:2008cx}, with some errors on the identification of global properties. Here, we take care of them in order to understand the vacuum structures. 

There are multiple crucial differences in the dynamics of the $K$ even and odd theories\footnote{In addition to the crucial differences discussed below, we need to take a different twisted boundary condition for fermions between odd and even $K$  as the symmetry structures are different. For odd $K$, we use (\ref{eq:chiral_symmetry_oddK}) to give real mass for fermions, but for even $K$, we need to use (\ref{eq:baryon_number_symmetry}). }. 
First,  the monopole operators are of the form \eqref{monopole} except that there is an irremovable $\theta$ angle. In the previous subsection,  we set  $\theta_1=\theta$ and the rest to zero. As a result, only the $j=1$ monopole vertices are modified:
\bea
{\cal M}_{1,i}(x)&&=\rme^{-S_I/N} \rme^{\im \bm{\alpha}_i\cdot \bm{\sigma}_{1}(x)}    (\bm{\alpha}_i\cdot {\bm \psi}_{K}) (\bm{\alpha}_i\cdot {\bm \psi}_{1})  \rme^{\im  \frac{\theta}{N}} ,  \quad (i=1,\ldots, N),  \cr \cr
{\cal M}_{j,i}(x)&&=\rme^{-S_I/N} \rme^{\im \bm{\alpha}_i\cdot \bm{\sigma}_{j}(x)}    (\bm{\alpha}_i\cdot {\bm \psi}_{j-1}) (\bm{\alpha}_i\cdot {\bm \psi}_{j}),     \quad (i=1,\ldots, N, \;\;  j=2, \ldots K) . 
\label{monopole-2}
\eea
Since bions in the center-symmetric background are topologically neutral, the theta angle dependence does not show up there, hence, at second order, we generate the same bion induced non-perturbative potential  $V^{(2)}$ 
\eqref{bion-pot} as in the $K$ odd case.   
 Furthermore, the minima of the potential is again given in \eqref{minima}. To count the ground states correctly, we need to remove all gauge redundancies, and at this step, there are some subtle differences compared to 
$K$ odd case. 
The  $  (\mathbb{Z}_{N})^{K-1} $ gauge identifications are 
\bea 
(\bm{\sigma}_{1},   \ldots, \bm{\sigma}_{K}) &\sim & (\bm{\sigma}_{1}, \ldots, \bm{\sigma}_{K}) \cr 
&&+   \frac{2 \pi }{N} {\bm \rho}  \Big[  (1, 0, -1, 0,\ldots, 0) l_1+ 
\ldots + (0,  \ldots, 0, 1,0-1, 0 ) l_{{K\over 2}-1} \Big] \qquad \qquad  \cr 
&&+  \frac{2 \pi }{N} {\bm \rho}  \Big[ (0,1, 0, -1, 0, \ldots, 0) m_1+ 
\ldots + (0,  \ldots, 0, 1,0-1) m_{{K\over 2}-1} \Big]  . 
\label{gaugeiden}
\eea 
Therefore, there are two gauge invariant  combinations   of the dual photon fields under the  $  (\mathbb{Z}_{N})^{K-1} $ gauge transformations. These are 
 \bea 
 \sum_{j\in {\rm odd}  } \bm{\sigma}_{j}(x), \qquad    \sum_{j\in {\rm even}  } \bm{\sigma}_{j}(x).  
 \label{gauge-invariant-2}
 \eea

These  combinations also arise naturally from two possible  types of polygonal monopole vertices, whose structure of fermionic zero modes coincides with that of the chiral order parameter on 
$\mathbb R^4$, given in \eqref{poligon-even}. These are 
 \bea 
{\cal {MR}}_{i, \rm odd} &=& \prod_{j \in {\rm odd}} {\cal M}_{j,i}  \sim  \rme^{- {K S_I\over 2N}+\im {\theta\over N}}   \rme^{\im  \bm{\alpha}_i\cdot  \left( \sum_{j\in {\rm odd}  } \bm{\sigma}_{j}(x) \right)}    \{\psi^{(i)}_2 \psi^{(i)}_3\} \cdots \{\psi^{(i)}_K \psi^{(i)}_1\}      ,   \\
 {\cal {MR}}_{i, \rm even}  &=&  \prod_{j \in {\rm even}}  {\cal M}_{j,i} \sim  \rme^{- {K S_I\over 2N}}   \rme^{\im  \bm{\alpha}_i\cdot  \left( \sum_{j\in {\rm even}  }^{K}  \bm{\sigma}_{j}(x) \right)}     \{\psi^{(i)}_1 \psi^{(i)}_2\} \cdots \{\psi^{(i)}_{K-1} \psi^{(i)}_K\} , 
\label{polygon}
\eea 
where $\psi^{(i)}_j=\bm{\alpha}_i\cdot \psi_j$. At $\theta=0$, these two polygonal vertices are related by $\mathbb{Z}_K$ permutation symmetry. 
Note that the fermionic zero mode structure of the two operators are the same up to contraction of spinor indices, but magnetic charges of the first (second) one lives on the odd  (even) sites,  and they are different.  Note that the exponents are the same as  the the gauge invariant combinations \eqref{gauge-invariant-2} as they should be.  Furthermore, there is a  $\theta$ angle dependence  in  one of them coming from irremovable $\theta$ angle in Lagrangian.

Under the  $\mathbb Z_{N}$  chiral transformation  \eqref{eq:chiral_symmetry_evenK},  the multi-fermion operators that enter to \eqref{polygon} transform by a  $\mathbb Z_{N}$  phase. Since $\mathbb Z_{N}$  is a genuine symmetry of the theory, this requires    
   \bea
\rme^{\im  \bm{\alpha}_i\cdot  \left( \sum_{j \in {\rm odd}}  \bm{\sigma}_{j}(x) \right)}  \rightarrow   \rme^{-{2\pi\im \over N}}  
\rme^{\im  \bm{\alpha}_i\cdot  \left( \sum_{j \in {\rm odd}}  \bm{\sigma}_{j}(x) \right)  } 
\cr  \cr
\rme^{\im  \bm{\alpha}_i\cdot  \left( \sum_{j \in {\rm even}}  \bm{\sigma}_{j}(x) \right)}  \rightarrow   \rme^{-{2\pi\im \over N}}  
\rme^{\im  \bm{\alpha}_i\cdot  \left( \sum_{j \in {\rm even}}  \bm{\sigma}_{j}(x) \right)  } 
 \eea 
Therefore, at the second-order in semi-classics, there are $N^2$ gauge inequivalent minima, given by
 \bea 
\sum_{j  \in {\rm odd}} \bm{\sigma}_{j}  =  \frac{2 \pi }{N} {\bm \rho} \ell_1 , \qquad  \ell_1 \equiv \Big( \sum_{j   \in {\rm odd}}  k_j  \Big) \;\;   {\rm mod} N  \cr 
\sum_{j  \in {\rm even}} \bm{\sigma}_{j}  =  \frac{2 \pi }{N} {\bm \rho} \ell_2 , \qquad  \ell_2 \equiv \Big( \sum_{j   \in {\rm even}}  k_j  \Big) \;\;   {\rm mod} N 
\label{gauge-invariant-3}
\eea 
 But this cannot be the number of vacua, as the chiral symmetry of the theory is $\mathbb Z_{N}$ and we cannot get more than $N$ vacua.  However, surprisingly, this number $N^2$ is physically meaningful.  

The first term that can lift the $N^2$-fold degeneracy appears at a fairly large  order in semi-classics. It may be induced by the correlated topological configurations of the form $[{\cal {MR}}_{i, \rm odd} \overline {\cal {MR}}_{i, \rm even}]$, which is a  $(2M)$-th  order event. This is an  exotic generalization of magnetic bion. The proliferation of such events induce a term in the effective theory of the form\footnote{While the sign of this contribution cannot be a priori established without careful semi-classical analysis, as physically meaningful signs could appear~\cite{Behtash:2015kna,Behtash:2015kva,Behtash:2015zha,Kozcaz:2016wvy}, we can fix the sign by demanding that the vacuum energy is minimal at $\theta=0\bmod 2\pi$. The justification of this can be argued by a version of the Vafa-Witten~\cite{Vafa:1983tf,Vafa:1984xg} theorem, combined with the continuity of all even-site quivers to the 2-site quivers discussed in sec.~\ref{sec:universality}. Since the even-site quiver is just a Dirac fermion coupled to $SU(N)\times SU(N)$ gauge field, the positivity of Euclidean measure is violated by the $\theta$-term only. Hence Vafa-Witten theorem applies, and the lowest vacuum energy is at $\theta=0\bmod 2\pi$.}:
\bea
V^{(2M)}&=&-\rme^{-KS_I/N} \sum_{i=1}^{N}  \cos \Big(  \bm{\alpha}_i \cdot  \Big( \sum_{j \in {\rm odd}}  \bm{\sigma}_{j}(x)  -  \sum_{j \in {\rm even}}  \bm{\sigma}_{j}(x) 
\Big) + \frac{\theta}{N}  \Big) 
\label{bion-pot}
\eea
As a result,  there are $N$ vacuum state for any value of  $\theta \neq \pi$.  For example, at $\theta=0$, we obtain the lifting term 
for the minima \eqref{gauge-invariant-3}, of the form $-\rme^{-(2M)S_I/N} N  \cos \left( \frac{2\pi}{N} (\ell_1 -\ell_2) \right)$,  
hence, the minima are at  $\ell_1 = \ell_2 =0, 1, \ldots N-1$.

Remarkably, since $\theta$ angle is physical in this theory, on each vacuum, we obtain $N$ branches as a function of the $\theta$-angle. 
This accounts for the $N^2$ minima that we found at second order in semi-classics.  Ultimately, $N^2$ minima   split up to $N$ vacua, each of which possess $N$  branches. 

The $\theta$ dependence of the vacuum energy  density also arises as
\be
E_{\ell_1,\ell_2}(\theta) \sim  \min_{(\ell_1-\ell_2)\in \mathbb{Z}}  \rme^{-KS_I/N}  N\left\{1-   \cos \Big(  \frac{ \theta + 2 \pi (\ell_1-\ell_2) }{N}  \Big)  \right\}. 
\ee
Here, we put an offset so that $E_{\ell_1=\ell_2}(\theta=0)=0$, so that $E_{\ell_1,\ell_2}(\theta)\ge 0$. 
This $\theta$ dependence is again a very suppressed  effect in semi-classics, but nonetheless  it is the leading term which  induces the branch structure. 
Notably, there is a first order phase transition at  $\theta=\pi$,  and number of vacua becomes   $2N$.

 On $\mathbb R^4$,  the order parameter for the chiral symmetry breaking is the   bosonic gauge invariant operator with $K$ fermion insertions,  $\mathcal{O}_{(1)}^{\mathrm{even}}=\tr([\psi_1\psi_2]\cdots[\psi_{K-1}\psi_K])$ and $\mathcal{O}_{(2)}^{\mathrm{even}}=\tr([\psi_2\psi_3]\cdots [\psi_K\psi_1])$.  The vacuum expectation value of $\mathcal{O}$ would yield, at $\theta \neq \pi$, to spontaneous  discrete chiral symmetry breaking  $\mathbb{Z}_{N} \rightarrow  \mathbb{Z}_{1} $, and $N$ vacua. However, due to physical theta angle in this theory, the condensate must have a non-trivial $\theta$ dependence, and we can see this explicitly in semi-classical regime.

 On small $\mathbb R^3 \times S^1$, the fermion zero mode structure of two types of operators (unlike $K$ odd case) \eqref{polygon} have exactly the right structure of zero modes to contribute to condensate. In the vacuum, the flux part of the monopole operator  given in \eqref{gauge-invariant-3} condense. Taking $\ell_1= \ell + \tilde \ell, \ell_2= \ell$,  the condensate has two types of contributions, 
\bea 
 \langle  \Omega_{\ell, \tilde \ell} | \left(\mathcal{O}_{(1)}^{\mathrm{even}}+\mathcal{O}_{(2)}^{\mathrm{even}} \right) | \Omega_{\ell, \tilde \ell}   \rangle  &\sim& N  L^{-3K/2} \rme^{ -  K \frac{8 \pi^2}{g^2 N}} \rme^{{2\pi\im   \ell  \over N}} \left( \rme^{  \im {{2\pi  \tilde   \ell  + \theta} \over N}}  +1 \right) \cr 
   &\sim&  N \Lambda^{3K/2} \rme^{{2\pi\im  \ell  \over N}} \left( \rme^{  \im {{2\pi  \tilde   \ell  + \theta} \over N}}  +1 \right) , \\
 \langle  \Omega_{\ell, \tilde \ell} | \left(\mathcal{O}_{(1)}^{\mathrm{even}}-\mathcal{O}_{(2)}^{\mathrm{even}} \right) | \Omega_{\ell, \tilde \ell}   \rangle &\sim & N \Lambda^{3K/2} \rme^{{2\pi\im  \ell  \over N}} \left( \rme^{  \im {{2\pi  \tilde   \ell  + \theta} \over N}}  -1 \right). 
\eea
The part of the condensate induced by ${\cal {MR}}_{i,\rm odd}$ has an explicit $\theta$ angle dependence, while the part  sourced by  ${\cal {MR}}_{i, \rm even} $ does not.  
At $\theta=0$, the second condensate vanishes for ground states $\tilde{\ell}=0$, which is the consequence of unbroken $\mathbb{Z}_K$ permutation symmetry. 
The $CP$ transformation acts on the chiral condensate as the complex conjugation. At the $CP$ symmetric points, $\theta=0$ or $\theta=\pi$, we can effectively write down its effect as the mapping of the labels $\ell,\tilde{\ell}$, but there is an important difference between $\theta=0$ and $\pi$. At $\theta=0$, the effect of $CP$ is 
\be
\ell\mapsto -\ell,\; \tilde{\ell}\mapsto -\tilde{\ell}. 
\ee
At $\theta=\pi$, however, the effect of $CP$ is given by 
\be
\ell\mapsto -\ell,\; \tilde{\ell}\mapsto -\tilde{\ell}-1.  
\ee
This shift on $\tilde{\ell}$ by $1$ is the consequence of mixed 't~Hooft anomaly or global inconsistency involving $CP$ at $\theta=\pi$, as we have discussed in the previous subsection. 

It is important to note that the magnitude of the condensate depends on the $\theta$ angle, unlike $K$ odd case,  as 
well as unlike general  $SU(N)$ QCD-like theories.  The magnitude of the condensate is: 

\bea 
\left|  \langle  \Omega_{\ell, \tilde \ell}  | \left(\mathcal{O}_{(1)}^{\mathrm{even}}+\mathcal{O}_{(2)}^{\mathrm{even}}\right) | \Omega_{\ell, \tilde \ell}   \rangle \right|   
   &\sim& \max_{\tilde \ell} N \Lambda^{3K/2}  \cos \Big({ {2\pi  \tilde   \ell  + \theta} \over 2N }  \Big) . 
\eea
Similar to $K$ odd case and  ${\cal N}=1$ SYM theory,  the $L$ dependence of the chiral condensate disappears within the one-loop $\beta$ function, and in a weakly coupled domain, we obtain a chiral condensate which is dictated by the strong scale $\Lambda$ of the theory. 
Note that the value of ${\tilde \ell}$ that minimizes the vacuum energy density maximizes the magnitude of the condensate. 

\section{Continuous deformation between quivers and universality of $K$-even and $K$-odd phases}\label{sec:universality}

Here we will argue that the vacuum structure of the even site quivers are continuously connected to each other, and the same holds for odd site quivers. 
This clarifies the reason why even-$K$ chiral quiver theories have the same properties with QCD(BF) and odd-$K$ theories do with $\mathcal{N}=1$ SYM. 

To argue this, we introduce a hierarchy of energy scales, so that one gauge theory of the quivers (say the one on site $i$) is more strongly coupled than the rest.
Then, in the deep UV regime, all gauge fields are weakly coupled, in the deep IR regime, all gauge fields are strongly coupled, but in the intermediate regime, the $i$-th gauge group is strongly coupled but the rest are weakly coupled. 
In this intermediate regime, the gauge group $SU(N)_i$ is expected to confine the fermions $\psi_{i-1}$ and $\psi_i$ into mesons and baryons. 
The color of the other gauge groups (in particular $i-1$ and  $i+1$) is weakly coupled at the intermediate scale, so a description in terms of gauge fields is still appropriate for them. 
The would-be pions of the site $i$ will then Higgs the gauge groups adjacent to site $i$ of the quiver down to a diagonal subgroup at the intermediate scale. The result is that the vertex $i$ of the quiver gaps out the fermions that connect to it, and the would-be pions in turn Higgs down the gauge groups $i\pm 1$ to a single $SU(N)$ gauge group: $SU(N)_{i-1}\times SU(N)_{i+1}\to SU(N)_{\mathrm{diag}}$. 
This effectively identifies gauge groups $i+1$ and $i-1$, and the $K$-site quiver is replaced with the $(K-2)$-site quiver. 

Now let us consider this procedure in a bit more details and take $\Lambda_{i}\gg\Lambda_{j\ne i}$ (Similar limit has been considered in the context of technicolor~\cite{Kaplan:1983sk, Georgi:1985hf, ArkaniHamed:2001ca}). 
In this case, the fermions $\psi_{i-1}$ and $\psi_i$ are strongly coupled to the gauge group $SU(N)_i$, while gauge fields at sites $i\pm 1$ are still weakly coupled in the intermediate range of energies. 
In this regime we expect that the gauge group  $i$ binds fermions into ``mesons'' $M_i\sim \langle \psi_{i-1}\psi_{i} \rangle$. The meson field $M_i$ is a nonlinear realization of Nambu-Goldstone bosons, $M_i\propto V_i \in SU(N)$, and it transforms as $V_i\mapsto U_{i-1} V_i U^\dagger_{i+1}$, under the $(i\pm 1)$-th gauge transformations $(U_{i-1},U_{i+1})\in SU(N)_{i-1}\times SU(N)_{i+1}$ .  
The leading term of an effective Lagrangian is given by
\be
\mathcal L\propto \Lambda^2_i\tr\left[{(\diff V_i+\im a_{i-1}V_i-\im V_i a_{i+1})^\dagger\wedge \star (\diff V_i+\im a_{i-1}V_i-\im V_i a_{i+1})}\right].
\ee
This is just the standard pion Lagrangian coupled to the $SU(N)\times SU(N)$ gauge fields.

In the limit that $\Lambda_i\gg \Lambda_{i\pm 1}$, we have that the above effective theory imposes a constraint
\be
a_{i+1}=V_i^\dagger a_{i-1}V_i-\im V_i^\dagger d V_{i}\;,
\ee
so the gauge fields $a_{i-1}$ and $a_{i+1}$ are forced to be the same up to a gauge transformation, and the gauge group $SU(N)_{i-1}\times SU(N)_{i+1}$ is reduced to a single $SU(N)$. Hence such a limit reduces the $K$-site quiver to the $K-2$ site quiver theory. In our discussion we have ignored the Wess-Zumino-Witten (WZW) term contribution \cite{Wess:1971yu, Witten:1983tw}. This term is important for the correct gauge anomaly cancellation. However as we have seen in this work, for $K>2$, the $K$- and $(K-2)$-site quiver gauge theories have the same 't Hooft anomalies, and the WZW term is not needed for anomaly matching. This is also justified by the observation that $V_i$ has no interesting dynamics as it can be gauged away to unity by a combined $i-1$ and $i+1$ gauge transformation. 

The discussion above implies that the $K$-site quiver theory can be reduced to the $(K-2)$-site quiver theory when $K>2$. 
That is, when $K$ is even, we can continuously deform our theory to QCD(BF), and when $K$ is odd, we obtain $\mathcal{N}=1$ SYM. 
This explains why these theories share the same structure about symmetry and 't~Hooft anomalies, as we have seen in previous sections. 
We conjecture that no phase transition is encountered as one takes the decoupling limit at a time. 
Even though nonperturbative orbifold equivalence does not hold for $K\ge 3$, there is an interesting continuity about ground-state structures between odd-$K$ theories, and the same holds, separately,  for even-$K$ theories. 

Lastly, let us comment on why we cannot apply the above procedure to QCD(BF) to obtain pure Yang-Mills theory, as we would naively conclude. 
Indeed, the situation is quite different for $K=2$ quiver with $\Lambda_2\gg \Lambda_1$, as there is only one weakly coupled gauge theory and the gauge transformation of $V=V_2$ is $V\rightarrow U V U^\dagger$. In this case $V$ transforms in the adjoint representation of $SU(N)$ gauge group, so the Nambu-Goldstone bosons cannot be gauged away completely. 
Depending on what value of $V$ is dynamically favored, the effective theory may be dynamically Abelianized and produce IR photons, or may be confining. 
If it is confining, the anomaly analysis dictates that the $\mathbb Z_N$ chiral symmetry is spontaneously broken, and indeed this was argued to be the case in Ref.~\cite{Karasik:2019bxn}.

\section{Comments on domain walls}\label{sec:domain_wall}

As we have discussed, chiral quiver gauge theories break discrete chiral symmetry spontaneously. In this case, there is a dynamical wall-type excitation, which connects different domains. 
Since such configuration breaks the spacetime translation, there is an associated Nambu-Goldstone modes on the wall. 
In addition to it, we must have extra light degrees of freedom because of the presence of various anomalies (see, e.g., Refs.~\cite{Gaiotto:2017yup, Gaiotto:2017tne, Komargodski:2017smk, Anber:2018jdf,Anber:2018xek, Nishimura:2019umw}), which makes domain-wall theories more interesting in our setup.

\subsection{Odd $K$ theories}

For odd $K$, we have  concluded that discrete chiral symmetry is spontaneously broken as $\mathbb{Z}_{2N}\to \mathbb{Z}_2$, and there are $N$ vacua. 
Let us consider the domain wall that connects two neighboring vacua. We assume that the domain wall locates at $x_3=0$, and 
\bea
&&\left.\langle\tr(\psi_1\cdots \psi_K \psi_1\cdots \psi_K)\rangle\right|_{x_3\to -\infty}=N\Lambda^{3K},\nonumber\\
&&\left.\langle\tr(\psi_1\cdots \psi_K \psi_1\cdots \psi_K)\rangle\right|_{x_3\to +\infty}=N\Lambda^{3K}\rme^{2\pi\im/N}.  
\label{DW}
\eea

The presence of anomaly (\ref{eq:anomaly_oddK}) suggests that, under a certain regularization, the bulk partition functions for $x_3\gtrless 0$ are given as 
\be
Z_{x_3<0}[B]=1,\; Z_{x_3>0}[B]=\exp\left(-\im\int_{x_3>0} {N\over 4\pi}B\wedge B\right). 
\ee
As a consequence, the partition function of the domain wall, $Z_{\mathrm{DW}}[B]$, must have an 't~Hooft anomaly for $\mathbb{Z}_N^{[1]}$ to cancel the anomaly inflow from the bulk partition functions:
\be
Z_{\mathrm{DW}}[B+\diff \lambda]=\exp\left(\im\int \left({N\over 2\pi}\lambda\wedge B+{N\over 4\pi}\lambda\wedge \diff \lambda\right)\right)Z_{\mathrm{DW}}[B]. 
\ee
A typical example having this anomaly is the level-$1$ $SU(N)$ Chern-Simons (CS) term.  
Although we cannot specify the domain-wall theory completely, consideration from anomaly suggests the existence of CS term in the low-energy effective theory\footnote{We are not claiming that the domain-wall theory is exactly given by the CS theory. It can include additional terms, which can change the details of dynamics. Still, the perimeter law of Wilson loops is robust under such deformation to satisfy the anomaly matching, so long as the domain-wall theory is gapped. }, and test quarks are deconfined on the wall (see also Refs.~\cite{Anber:2015kea, Sulejmanpasic:2016uwq}).

An intuitive way to see the deconfinement of test charges  is as follows. 
Let us start with the case $K=1$, corresponding  to ${\cal N}=1$ SYM, which is relatively  well understood with the help of $\mathcal{N}=2$ Seiberg-Witten (SW) theory~\cite{Seiberg:1994rs, Seiberg:1994aj}.  
In the moduli space of Seiberg-Witten theory, there are special points at which either the 't~Hoof-Polyakov monopole or Julia-Zee dyon become massless. 
By adding the soft mass to adjoint scalar, it is widely believed that these massless magnetically charged particles condense, which leads to confinement. 
Different $N$ vacua of $\mathcal{N}=1$ SYM theory can then be associated with the monopole-condensation and dyon-condensation phases. 
We note that semiclassical quantization with dyons fixes their possible electric charges~\cite{Tomboulis:1975qt}: They should belong to the root lattice, not the weight lattice, so their electromagnetic charges can be thought of the same with composites of one 't~Hooft-Polyakov monopole and multiple gluons. 

In this interpretation, the domain wall for spontaneously-broken discrete chiral symmetry is the wall configuration separating the monopole-condensation phase and the dyon-condensation phase\footnote{Note that this explanation differs from the one by Witten~\cite{Witten:1997ep}  (often attributed to S.~J.~Rey's unpublished work), although they may sound similar. 
In their explanation, condensing dyons have the electric charges in the weight lattice, i.e. they can take the fundamental representation, which causes the dynamical excitation on the wall with fundamental representation. 
In the explanation here, all dynamical dyons have the adjoint representation: What is argued here is the deconfinement of test electric charges on the wall, and there are no dynamical fundamental excitations. 
}. 
Because of the condensation of magnetically-charged particles, the dual-superconductor scenario~\cite{Nambu:1974zg, Mandelstam:1974pi, tHooft:1981bkw} naturally expects the area law of Wilson loops,  
\begin{align}
W^q,  \qquad  q=1, \ldots, N-1, 
\end{align}
 on both sides of the wall. 
Near the wall, there are several possibilities, and one possibility is the following. Because monopole and dyon have non-parallel charges, their condensates should decrease near the wall, which leads to the proliferation of electrically-charged excitations. 
Since those excitations are in the adjoint representation, it is natural to expect the screening of test electric particles. 
This is consistent with the deconfinement phenomenon for the $3$d CS theory. 

For $K \geq  3$, confinement can be tested by Wilson operators of the form $W_1^{q_1}  W_2^{q_2}  \ldots  W_K^{q_K}$. But since the massless bi-fundamental quarks $\psi_j$ has $N$-ality $(0, \ldots, +1, -1, 0, \ldots, 0)$ where $+1$ appears on $j$-th spot,  their pair-creation/annihilation processes  can change $(q_1, \ldots, q_K)$ 
without energetic cost according to the rule:
\bea 
(q_{1}, \ldots, q_{K}) \sim  (q_{1}, \ldots, q_{K})  +    (1, -1, \ldots, 0) l_1+ 
\ldots + (0,  \ldots, 0, 1,-1 ) l_{K-1}, \qquad \qquad 
\label{identify}
\eea 
with $l_i\in \mathbb{Z}$. 
 Therefore, external probes can be classified according to a single $\mathbb Z_N$ valued integer,  
 \bea 
\sum_{j=1}^{K}  q_j  =  q \bmod N . 
\label{classes}
\eea 
Without loss of generality, we can consider  external probes of the form $(0, q, 0, \ldots, 0)$ corresponding to 
\begin{align}
W_2^q,  \qquad  q=1, \ldots, N-1 
\end{align}
just like external probes in $K=1$ theory, corresponding to ${\cal N}=1$ SYM theory.  (We chose a Wilson line in   $j=2$ for later convenience).  
So, the classification of the external probes are, not surprisingly, 
same in $K=1$ vector-like theory and $K \geq 3$ chiral theories, since in both cases, we only have a single  $\mathbb Z_N^{[1]}$ center symmetry.

The above  explanation implies that it suffices to generate CS term only for one gauge factor. Let us show this explicitly. 
Consider the $x_3$ direction (direction transverse to DW) compactitfied on a circle with arbitrary size $\beta$.   Then, the insertion of domain wall  \eqref{DW} correspond to the chirally twisted boundary conditions in terms of microscopic fermions  in the path integral formulation, namely, 
\be 
  \psi_1 (\beta)  =  \rme^{2\pi\im \over 2N} \psi_1(0), \quad  \psi_j (\beta)= \rme^{{2\pi\im\over 2N} (-1)^j}\psi_j(0),\,\, (j=2,\ldots,K). 
 \ee
Ref. \cite{Poppitz:2008hr} showed that this chiral twist in the small $\beta$ regime induce CS terms, $CS_j$, for $SU(N)_j$. In the present case, integrating  over 
  $\psi_1$ induces $\frac{1}{2}(CS_1+ CS_2)$ and integration over   $\psi_j$ induces $ \frac{1}{2} (-1)^j (CS_j+ CS_{j+1})$, with $j=2, \ldots, K $. 
  After pairwise cancellations, the combined effect is just level-1 $CS_2$ with action 
 \be 
\Delta S= \frac{1}{4\pi} \int_{\mathbb R^3} \left ( a_2\diff a_2 + \frac{2\im}{3}a_2^3 \right), 
 \ee
which is indeed capable to produce deconfinement for $W_2^q$ on the wall.  In this set-up, since $S^1$  circle size is reduced  to a small value while keeping boundary conditions chirally twisted,  the theory on small  $\mathbb R^3  \times S^1$ can be interpreted as the world-volume theory of the domain wall, 
which include apart other things a Chern-Simons term.  

Finally, note that one can also consider one more compactification. Let us denote the world-volume directions of the DW as $(x_1, x_2, x_4)$ and the transverse direction as $x_3$. Let us assume the $x_3$ direction is non-compact, as in our original set-up \eqref{DW}. Let us now compactify 
$x_4$  to a small-circle and impose the double-trace deformation to have abelianization.   Then, the domain wall becomes a domain line with coordinates $(x_1, x_2)$, and the bulk is described in terms of coordinates  $(x_1, x_2, x_3)$ with a domain line located somewhere on $x_3$ coordinate. 
 We can impose  
 \bea
&&\left \langle \rme^{\im  \bm{\alpha}_i\cdot  \left( \sum_{j=1}^{K}  \bm{\sigma}_{j}(x) \right)}   \right \rangle    \Big |_{x_3 \rightarrow -\infty } = 1  \cr
&&\left \langle \rme^{\im  \bm{\alpha}_i\cdot  \left( \sum_{j=1}^{K}  \bm{\sigma}_{j}(x) \right)}   \right \rangle    \Big |_{x_3 \rightarrow +\infty } =
\rme^{{2\pi\im  \over N}} 
\label{condensate-smallL-wall}
 \eea
corresponding to different vacua in the semi-classical domain, and a domain line separating them.  Note that \eqref{condensate-smallL-wall} implies 
\eqref{DW} in semi-classical domain. 
In this case, the domain line theory 
includes the $BF$-type topological field theory that leads to deconfinement, as discussed in Ref.~\cite{Anber:2015kea, Cox:2019aji} for SYM.  It seems plausible that this set-up on small $S^1 \times \mathbb R^3$ is continuously connected to the  \eqref{DW} set-up on $\mathbb R^4$.

\subsection{Even $K$ theories}

When $K$ is even, the symmetry breaking patters are different for $\theta\not=\pi$ and $\theta=\pi$. For generic values of $\theta(\not=\pi)$, we have the discrete chiral symmetry breaking $\mathbb{Z}_N\to 1$, and for $\theta=\pi$, we also have the spontaneous $CP$ breaking, $\mathbb{Z}_N\rtimes (\mathbb{Z}_2)_{CP}\to 1$. 
In this section, we concentrate on the domain wall connecting two neighboring vacua by discrete chiral symmetry breaking. 

Because of the anomaly (\ref{eq:anomaly_oddK}) of the bulk, the domain-wall theory must have the following anomaly:
\be
Z_{\mathrm{DW}}[B+\diff \lambda, A_\mathrm{B}+\diff \phi]=\exp\left(\im\int {1\over 2\pi}\lambda \wedge (2N B+N\diff \lambda+\diff A_\mathrm{B})\right)Z_{\mathrm{DW}}[B, A_\mathrm{B}]. 
\ee
The first two terms represent the anomaly of $\mathbb{Z}_N^{[1]}$, which can be matched by level-$2$ $SU(N)$ Chern-Simons term. The last term implies the mixed 't~Hooft anomaly between $\mathbb{Z}_N^{[1]}$ and $U(1)/\mathbb{Z}_N$.  
Assuming the complete deconfinement, both anomaly can be matched at the same time. 
Therefore, deconfinement of test charges with any nontrivial $N$-ality is one of the consistent scenarios for the dynamics on the wall. 

While the detailed studies of the domain wall theory are an interesting topic, we will defer it to the future. Still it may be useful to discuss several proposals of theories which saturate the anomaly.

Consider a $U(1)_{2N}$ CS theory, with the Lagrangian
\be
\frac{\im N}{2\pi}a\wedge \diff a\;.
\ee
The theory has a $\mathbb Z_{2N}^{[1]}$ symmetry. By gauging the subgroup $\mathbb Z_{N}^{[1]}$ we easily see that the model has the correct anomaly $\im \frac{1}{2\pi}\int B\wedge B$. However the model has two deficits: 1. it has a larger one-form symmetry group and 2. it has no $U(1)$ symmetry which can act as the baryon symmetry.

Both of these issues are resolved by introducing matter field $\phi$ which is charged with an $N$-charge under the $U(1)$ gauge symmetry. Now this immediately reduces $\mathbb Z_{2N}^{[1]}\rightarrow \mathbb Z_{N}^{[1]}$. Moreover it allows a new symmetry -- topological -- with a conserved current $j=\frac{1}{2\pi}F$, where $F=\diff a$ is the curvature of $a$. Note that this was not a symmetry without the matter field $\phi$, as there was no gauge-invariant local operator which transforms under it. Indeed a monopole operator $M(x)$ is not gauge invariant because of the Chern-Simons term, and thus transforms as $M(x)\rightarrow M(x)\rme^{\im 2N\alpha(x)}$~\cite{Lee:1991ge, Pisarski:1986gr, PhysRevLett.66.276, Affleck:1989qf}. 
Under the presence of the charge $N$ matter field $\phi(x)$, however, the operator $M(x)\phi^*(x)^2$ is gauge invariant, and it transforms under the $U(1)$ topological symmetry. It is this symmetry that we want to associate with the $U(1)$ baryon-number symmetry. 

Gauging the $U(1)$ baryon symmetry, we have a term
\be
\frac{\im}{2\pi}\int_3 A_\mathrm{B}\wedge F=\frac{\im}{2\pi}\int_4 \diff A_\mathrm{B}\wedge F\;,
\ee
where in the second step we wrote the term in terms of an auxiliary dimension, which we are free to do.
Now, when we gauge the center symmetry, we further replace $F\rightarrow F+B$ and obtain the correct anomaly.

Let us discuss some limits of this model. First, we choose $\phi$ to be a very massive scalar. In this case, the model is gapped and has topological order, as the massive, gapped excitations $\phi$ obey anyonic statistics due to the CS term. In this phase, the $U(1)_B$ symmetry is not spontaneously broken as the CS term causes the expectation value of the monopole operator to vanish. In this phase, the $\phi$ excitation carries $1/2$ the baryon number.

On the other hand, we can take the opposite limit where the mass-squared of the scalar is taken to be negative, with a large absolute  value. Then, the gauge field $a$ will get Higgsed down to a $\mathbb{Z}_N$ gauge field, which provides a total-deconfinement scenario. 
Such a theory can be described as a $BF$ theory with a term $\frac{\im N}{2\pi}\int a\wedge \diff b$, where $b$ is a $U(1)$ gauge field to impose the constraint that $a$ is a $\mathbb Z_N$ gauge field. Note that the theory supports a $1/N$ fractional gauge-vortex which carries a $1/N$ baryon number.

The model above can also be obtained as a limit of other models, such as having two $U(1)_N$ CS theories, perhaps associated with the $\mathbb Z_2\subset \mathbb Z_K$ symmetry, and a Higgs field which forces them to be the same. Another proposal is a $U(2)_N$ CS theory, which can again be reduced to the above proposal by Higgsing it down, and postulating appropriate matter so that monopole operators can be gauge invariant.

\section{Discussions}

In this paper, we have discussed the dynamics of chiral quiver gauge theories. Despite the fact that they are chiral theories, their dynamics show surprisingly similar behaviors with those of vector-like gauge theories. 
For odd-sites quiver theories, the symmetry realizations, including anomalies, are the same with those of $\mathcal{N}=1$ SYM, and we have $N$ vacua by spontaneous discrete chiral symmetry breaking, $\mathbb{Z}_{2N}\to \mathbb{Z}_2$. 
Moreover, we have argued that all odd-sites quiver theories belong to the same phase and can be continuously connected by taking individual gauge theories to be much more strongly coupled than one at a time.
The even-sites quivers can all be continuously deformed to QCD(BF), which has $N$ vacua due to the chiral symmetry breaking, $\mathbb{Z}_N\to 1$, at generic $\theta$ angles, and has $2N$ vacua at $\theta=\pi$ due to the extra $CP$ breaking.
Reliable semiclassics on $\mathbb{R}^3\times S^1$ provides the concrete realization of these dynamics in a calculable regime, which brings us the better understanding of these anomaly constraints. 

In the case of even-sites quivers, $N=2$ may be special as we have briefly mentioned in the last paragraph of Sec.~\ref{sec:anomaly_even_sites}. 
Since there is no center-chiral mixed anomaly, symmetry breaking of the baryon number, $U(1)/\mathbb{Z}_N$, is also a viable scenario for low-energy dynamics. 
Since this scenario does not occur in our semiclassical analysis, we are inclined to think that it is unlikely. Nevertheless it is possible that the continuity to the theory on $\mathbb R^3\times S^1$ fails.
It would be very interesting if some future studies could elucidates the special or non-special nature for even-sites $SU(2)$ quiver gauge theories. 

Another open problem in this paper is regarding the theory on the domain walls. Thanks to the presence of the anomaly, as we have discussed in Sec.~\ref{sec:domain_wall}, the domain wall must support nontrivial low-energy effective theories. 
We provided some scenarios for them from the viewpoint of anomaly inflow, but we have not discussed its dynamics directly. 
It is perhaps notable that amongst the possible scenarios discussed here we have found both a scenario where a domain wall supports gapped $1/N$ baryon number excitations, as well as $1/2$ baryon excitations. The anomalies alone do not appear sufficient to fully constraint the phenomenology of the domain wall.
It would be very interesting if a better understanding of the domain walls can be established in the future.

\acknowledgments
The authors thank Mohamed Anber for useful comments on the draft and discussions. 
The work of T.~S. is funded by the Royal Society University Research Fellowship.
The work of Y.~T. was supported by the JSPS Overseas Research Fellowship until March 2020, and is supported by Yukawa Institute for Theoretical Physics.
The work of M.~\"{U}. is supported by the U.S. Department of Energy, Office of Science, Division of Nuclear Physics under Award DE-SC0013036.

\bibliographystyle{utphys}
\bibliography{./QFT}
\end{document}